\documentclass{aa}  

\usepackage{graphicx}
\usepackage{pdflscape}
\usepackage{color}
\usepackage{gensymb}
\usepackage{txfonts}
\usepackage{natbib}
\bibpunct{(}{)}{;}{a}{}{,} 

\begin{document}

\title{A KMOS survey of the nuclear disk of the Milky Way I: Survey design and metallicities \thanks{Based on ESO program 0101.B-0354.}} 

\subtitle{}

   \author{T. K. Fritz\inst{1,2}
   \and
             L. R. Patrick\inst{1,2,10}
                   \and            
                   A. Feldmeier-Krause \inst{6,5}
                   \and
                             R. Sch\"odel\inst{3}          
          \and 
          M. Schultheis \inst{7}
                             \and 
                             O. Gerhard\inst{4}
                                       \and 
          G. Nandakumar \inst{8,9} 
          \and 
          N. Neumayer\inst{5}
          \and 
          F. Nogueras-Lara \inst{5} 
            \and
            M. A. Prieto\inst{1,2}
}   

              \institute{Instituto de Astrofisica de Canarias, calle Via Lactea s/n, E-38205 La Laguna, Tenerife, Spain\\
               \email{tobias.k.fritz@gmail.com}
          \and
          Universidad de La Laguna, Dpto. Astrofisica, E-38206 La Laguna, Tenerife, Spain   
        \and   Instituto de Astrofisica de Andalucia (CSIC), Glorieta de la Astronomia s/n, 18008 Granada, Spain
             \and  Max Planck Institute for extraterrestrial Physics, Giessenbachstraße 1, D-85748 Garching, Germany
             \and Max-Planck-Institut für Astronomie, Königstuhl 17, 69117, Heidelberg, Germany
              \and The Department of Astronomy and Astrophysics, The University of Chicago, 5640 S. Ellis Ave., Chicago, IL, 60637, USA
                \and Laboratoire Lagrange, Université Côte d'Azur, Observatoire de la Côte d'Azur, CNRS, Blvd de l'Observatoire, 06304, Nice
                \and Research School of Astronomy \& Astrophysics, Australian National University, ACT 2611, Australia 
                \and ARC Centre of Excellence for All Sky Astrophysics in Three Dimensions (ASTRO-3D), Australia 
                  \and Departamento de F\'{\i}sica Aplicada, Facultad de Ciencias, Universidad de Alicante, Carretera San Vicente s/n, E-03690, San Vicente del Raspeig, Spain
             }

   \date{}

 
  \abstract{
  {In the central few degrees of the bulge of the Milky Way there is a flattened structure of gas, dust and stars (the central molecular zone) similar to nuclear disks in other galaxies.
  Due to extreme foreground extinction we possess only sparse information about the (mostly old) stellar population of the nuclear disc.  
}
{Here we  present our KMOS spectroscopic survey of the stars in the nuclear disk reaching the old populations. 
In order to obtain an unbiased data set, we sample stars in the full extinction range along each line-of-sight.
}
{We also observe reference fields in neighboring regions of the Galactic bulge. Here we describe the design and execution of the survey and present first results.}
{We obtain spectra and five spectral indices of 3113 stars with a median S/N of 67 and measure radial velocities for 3051 stars. 
Of those, 2735 sources have sufficient S/N to estimate temperatures and metallicities from indices. 
}
{ %
We derive metallicities using 
the CO 2-0 and Na I $K$-band spectral features, where we derive our own empirical calibration using metallicities obtained with higher resolution observations.
We use 183 giant stars for calibration spanning in metallicity from -2.5 to 0.6 dex and covering temperatures of up to 5500 K.
The derived index based metallicities deviate from the calibration values with a scatter of 0.32 dex. The internal uncertainty of our metallicities is likely smaller. 
 We use these metallicity measurements together with the CO index to derive effective temperatures using literature relations. We publish the catalog here. Our data set complements Galactic surveys such as \textit{Gaia} and APOGEE for the inner 200 pc radius of the Milky Way 
which is not readily accessible by those surveys due to extinction.
We will use the derived properties in future papers for further analysis of the nuclear disk. 
}
  } 

\titlerunning{A KMOS survey of the nuclear disk of the Milky Way I}
\authorrunning{T. Fritz et al.}   

\maketitle
%
\section{Introduction}
\label{sec:intro}
 The nuclear regions of galaxies show morphological and kinematic structures on scales of a parsec to a some hundred parsecs that clearly set them apart from the large-scale structures of galaxies. 
 Around the central black holes \citep{Kormendy_13} there are additional structures in stars, dust and gas.
  On the scale of a few parsecs there are compact  nuclear star clusters from $10^5$ to a few $10^7$ M$_\odot$ \citep{Boeker_02,Walcher_05,Chatzopoulos_15,Neumayer_20} in almost every galaxy with mass similar to the one of the Milky Way.
  Further out, in spiral galaxies, particularly in barred spiral galaxies, there are more flattened structures 
  like nuclear rings, inner bars and disks \citep{Erwin_02,Launhardt_02,Comeron_10,Gadotti_19} of about a few 100 pc and up to 1 kpc \citep{Gadotti_20}\footnote{For simplicity we will call all nuclear structures nuclear disks in the following.}. Often there is more 
  dust and late star formation in the nuclear region than in the surrounding bulge and inner disk \citep{Gadotti_19,Bittner_20}.
 
 Nuclear components are interesting on their own as well as the way they are connected to the larger scales of a galaxy.
  Galactic centers are sinks of gas. In particular, the bar moves gas towards galactic centers \citep{Athanassoula_83,Binney_91,Emsellem_15}. It accumulates at the inner Lindblad resonance of the bar and forms stars there \citep{Athanassoula_83,Knapen_05,Kim_12}, often in rings. While the gas that is currently observed there, was probably only recently deposited, the stars keep an imprint of gas inflows that happened in the past.
  The star formation history can be used to infer how efficiently the bar transported gas to the center over time \citep{Nogueras_19c,Bittner_20}. 
  
    Further, a nuclear star formation burst may influence the outer galaxy through outflows \citep{Veilleux_05} and form structures like the Fermi bubble \citep{Su_10}, the radio bubble \citep{Heywood_19} and the X-ray chimney \citep{Ponti_19} in the Milky Way. 
  In the other direction, compact structures like globular clusters can migrate towards the Galactic center due to dynamic friction \citep{Tremaine_75,Tsatsi_17} and deposit stars there. Such stars can be identified from their low metallicites \citep{Dong_17}. 
  It is also possible that the nuclear cluster stars were formed locally or in locally formed clusters \citep{Milosavljevic_04,Mastrobuono-Battisti_19}.
  
    All three nuclear components (central black hole \citep{Schoedel_02}, nuclear cluster \citep{Becklin_68,Catchpole_90,Launhardt_02}, and nuclear disk) exist together in the Milky Way, embedded into the  bulge. In contrast to the bulge, there are young stars in clusters \citep{Figer_99} and outside \citep{Cotera_96,Cotera_99,Clark_21} in the nuclear disk. There is also molecular gas, the central molecular zone \citep[CMZ][]{Morris_96,Mezger_96,Molinari_11,Kruijssen_15,Ginsburg_16}, with about the same extent as the stars of the nuclear disk \citep{Launhardt_02}.
    The nuclear cluster of the Milky Way has a size of about 5 pc \citep{Fritz_16,Gallego_20}.
    The nuclear disk has a radius of about 230 pc and a scale height of about 45 pc \citep{Launhardt_02}, similar but somewhat smaller than most extragalactic nuclear structures \citep{Gadotti_19,Gadotti_20}. 
  
  Comparing the chemistry of the oldest stars in the nuclear disk and the inner Galactic Bulge \citep{Ness_13,Schultheis_15, Zoccali_17}
  will show whether they are chemically similar.
  In that case the old nuclear disk stars are possibly (partly) just the inner continuation of the bulge, otherwise the stars are possibly connected to precursors of today's CMZ. There is an indication for differences between the nuclear cluster and the bulge \citep{Nogueras_18b,Schultheis_19,Schoedel_20}. 
  The  bulge of the Milky Way is in its outer parts bar-shaped and mostly formed by secular evolution \citep{Bland_16}.
  However, it is still possible that an old classical bulge is hidden in the inner  metal-poor bulge  \citep{Dekany_13,Kunder_20,Arentsen_20}. The coverage of the inner bulge is still too limited to answer this question.

  Photometric data of the nuclear disk are available from large scale surveys. The most extensive bulge survey is VVV \citep{Minniti_10}. SIRIUS \citep{Nishiyama_13} covers the full nuclear disk and parts of the surrounding  bulge. GALACTICNUCLEUS \citep{Nogueras_19a} also covers the central part of the nuclear disk at larger depth and sensitivity.  The nuclear disc is much too extincted for \textit{Gaia} proper motions\citep{Brown_18}. HST proper motions are now available in a relatively small area \citep{Libralato_21}, while crowding and saturation makes the larger area effort by VIRAC \citep{Clarke_19} less useful. Due to the large and variable extinction \citep{Schoedel_10,Fritz_11,Nogueras_18}, metallicity estimates based on color magnitude diagrams are difficult to obtain
  \citep{Gonzalez_13}, even J-band images are difficult to obtain at sufficient depth.
  
    While the central few parsecs of the Milky Way have been already observed extensively with spectroscopic observations and high resolution imaging \citep{Genzel_10}, 
    the nuclear disk has not yet been studied in depth \citep{Bland_16}.  
The largest scale work is radio based, consists of 
Masers \citep{Lindqvist_92,Trapp_18} and thus covers only a short and not well understood evolutionary phase. 
Outside the few known clusters, the nuclear cluster \citep{Feldmeier_14,Do_15,Ryde_16,Fritz_16,Do_18,Thorsbro_20,Feldmeier_20,Davidge_20}, 
Arches and Quintuplet \citep{Najarro_04,Martins_08,Clark_18a,Clark_18b}, near infrared spectroscopy is limited, and was mostly aimed to target special stars like young stellar objects \citep{Nandakumar_18}, early-type stars with strong emission lines and X-ray activity \citep{Mauerhan_10a,Mauerhan_10c,Clark_21} 
or very red stars \citep{Geballe_19}. The most extensive data till now are likely from APOGEE \citep{Schoenrich_15,Majewski_17,Schultheis_20}.
However, APOGEE is not ideal to study the Galactic nuclear disk: the relatively small telescope size and high extinction in the H-band limit the APOGEE observations, and bias the sample to blue and intrinsically bright stars.
  
 Therefore we executed a dedicated spectroscopic survey of the central 270 $\times$ 130 pc  radius of the Milky Way targeting the nuclear disk and the innermost bulge in the IR-K-band using KMOS (VLT).  
This paper is the first of a series. The
paper is structured as follows: in Sect.~\ref{sect:surv_des} we describe our survey design and the obtained data.
We detail the data reduction procedures in Sect.~\ref{sect:data} including extraction of 1D spectra of the targets.  We analyze the stellar spectra in Sect.~\ref{sec:spec_ana}, and measure the line-of-sight velocity and line index values. We use the latter to derive metallicities and effective temperatures in Sect.~\ref{sec:phys_properties}. That includes a new calibration for deriving metallicities from line indices. 
We give summary and conclusions in Sect.~\ref{sec:summ}.


\section{Survey design}
\label{sect:surv_des}

Here we describe the design of our survey. First, we describe the general survey properties, and how we select the potential targets and fields for our KMOS observations. Finally, we describe the actually targeted stars and the observation setup. 

\subsection{General survey properties}
\label{sect:surv_gen}

Our dedicated survey of the nuclear disk has the following properties:

\begin{itemize}
\item In order to characterize the full nuclear disk, our survey samples the full range of the nuclear disk
 \citep[radius of $\approx1.55$\degree\,$\approx$220 pc and scale-height of $\approx0.3$\degree\,$\approx$45 pc, see][]{Launhardt_02} 
and line-of-sight depth.
\item  Due to the relatively small size \citep{Launhardt_02,Nishiyama_13,Gallego_20} the line-of-sight distances of nuclear disc stars vary little, but there is a significant amount of dust in the nuclear region \citep{Launhardt_02,Chatzopoulos_15a,Nogueras_20}, 
that obscures stars at varying depth. For this reason, we selected our targets in extinction corrected magnitudes.
\item Our survey is designed to cover the full range in age and metallicity. The population with
the faintest tip of the red giant branch (RGB) is old and metal-poor. Theoretical PARSEC isochrones \citep{Bressan_12,Marigo_17} predict that the tip of the RGB of an old, metal-poor stellar population (12 Gyr, [Fe/H]=-1)  is
at M$_H=-6.09$ and M$_K=-6.25$. In order to reach the old stars of the nuclear disk, we selected predominantly stars fainter than this limit. 
The magnitude range of our sample extends over more than two magnitudes.
Because of this, a small change in the extinction law over our fields does not affect the selected sample significantly. 
Further, metal-poor stars are bluer in (H-K) color. To prevent biases, we carefully selected stars distributed over the full range of (H-K) color in the nuclear disk. 
\end{itemize}

We measure line-of-sight velocities of individual stars for dynamics. 
A relatively low resolution is sufficient to measure the internal dispersion of about 70 km/s.
To obtain the star formation history, we construct a Hertzsprung-Russell diagram with luminosities and temperatures. Due to the high extinction, temperatures need to be derived from spectra rather than from photometry. 
We also determine metallicities for most stars.
Because the measurement of absolute properties, especially metallicities, is difficult, we obtain the same kind of observations also outside the nuclear disk in the inner bulge. That way we can account for bulge pollution in our nuclear disk sample. Since the bulge is probably symmetric in star formation history and chemistry in latitude \citep{Nandakumar_18b}, sampling it on one side is sufficient. 

The interstellar extinction toward the Galactic center is high (up to A$_K=$4.5, A$_H=$8 even when excluding infrared dark clouds).
  The magnitude of the tip of the RGB of old metal poor stars and distance to the Galactic center of 8.18 kpc  \citep{Abuter_19} implies that stars fainter than m$_K=12.8$ or m$_H=16.5$ should be observed in a nuclear disk survey to sample the RGB tip completely.
This H-band magnitude is far below the limit of APOGEE \citep{Majewski_17} and is even challenging with H-band instruments at larger telescopes when more information than the line-of-sight velocity is wanted. 
Therefore, we observe exclusively in the K-band.

\subsection{Observed fields}
\label{sect:surv_kmos}

   \begin{figure*}
   \centering
   \includegraphics[width=0.76\columnwidth,angle=-90]{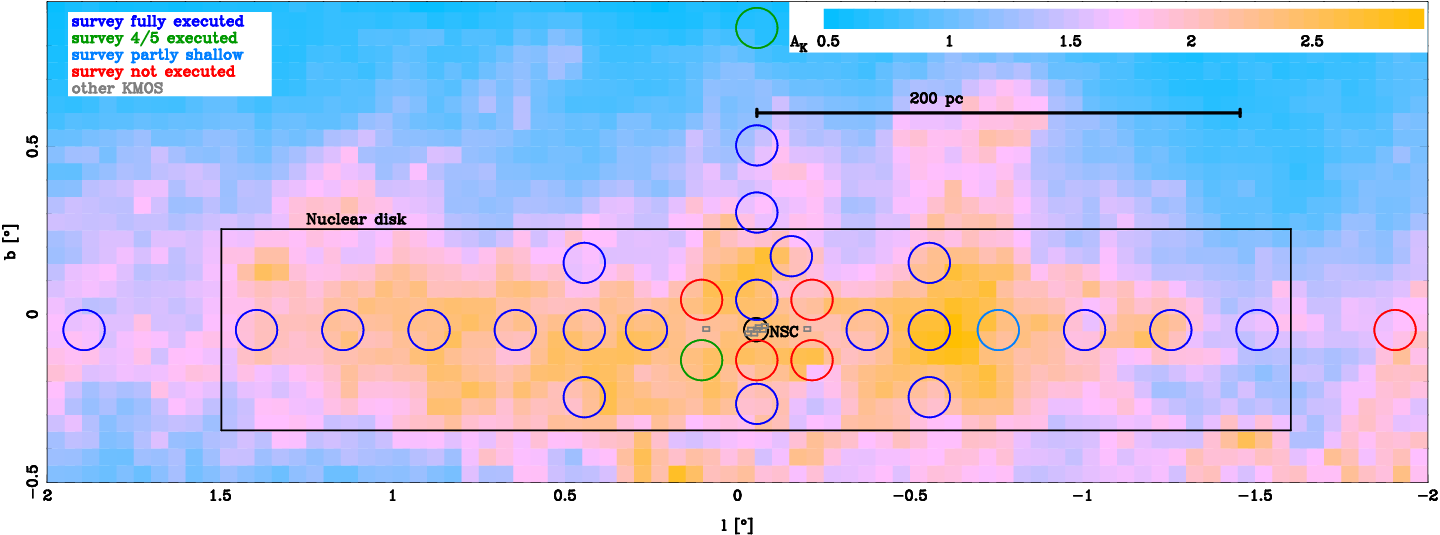}     
      \caption{Locations of the KMOS patrol fields of our survey. As shown by the colors most were executed. We also show similar KMOS observations outside of this survey (\citep{Feldmeier_20}; Feldmeier-Krause et al. in prep.). Together the full nuclear disk \citep{Launhardt_02} is covered. The small black circle in the center shows the effective radius of the nuclear cluster \citep{Gallego_20}.
      The background shows a low resolution extinction map in which the average extinction varies between A$_{K}=0.45$ and 2.93, see color bar. We made the extinction map from the catalog of \citet{Nishiyama_13}. 
              }
         \label{fig:field_loc}
   \end{figure*}
 
We observe with the instrument K-band Multi Object Spectrograph (KMOS) \citep{Sharples_13} at the VLT.
KMOS places 24 integral field units (IFUs) of 2.8\arcsec$\times$2.8\arcsec size over a diameter of 7.2\arcmin.
The survey was executed in program 0101.B-0354 between April and September 2018.
With the survey we target 24 fields in the nuclear disk and five reference fields in the bulge.  
Two bulge fields are located on both sides of the nucleus in the plane and three at increasing Galactic latitude at the same Galactic longitude as Sgr~A*.  We tried to sample the nuclear region evenly.
We note that we avoided the inner few arcmin (the nuclear cluster) as it has been covered already by other programs \citep{Fritz_16,Feldmeier_17a,Feldmeier_20}. Not all requested fields in the central 0.2\degree~region were observed (red circles in Fig.~\ref{fig:field_loc}).
In the central part we also placed some fields outside the midplane to be able to measure properties and gradients in scale-height. In the outer parts where the nuclear disk is likely thinner \citep{Launhardt_02} we only target it directly in the Galactic plane.

\subsection{Target selection}
\label{sect:targ_select}

We use 24 IFUs of KMOS for target observations.
As targets we select stars primarily by $K$-band extinction corrected magnitude (K$_0$). We target stars with  K$_0$ between 7 and 9.5. For a Galactic center distance of 8.18 kpc \citep{Abuter_19} this corresponds to absolute magnitudes between -7.56 and -5.06. 
Thus, metal-poor old stars still fall within the selection window. 
Observationally, due to the  extinction range from about A$_{K}=0.5$ (in the outermost bulge reference field) to about 4.5, that corresponds to observed magnitudes of m$_{K}=7.5$ to 15.
As the primary catalog we use SIRIUS \citep{Nagayama_03,Nishiyama_06}. Compared to VVV \citep{Minniti_10} and 2MASS \citep{Skrutskie_06}, SIRIUS has the advantage that neither saturation nor crowding affects the majority of our targets. 
We use VVV photometry where SIRIUS is not available.
This affects two fields, in which about 12\% are missing. 
Stars brighter than $K = 9.2$ are saturated in SIRIUS, see also \citet{Matsunaga_09}; for them we use 2MASS instead.
This affects on average 3\% of targets per field, but for the outer most bulge field, this rises to 30\%.  
We align the 2MASS/VVV magnitude to SIRIUS by using the median of common sources to ensure that all stars are on the same calibration.
We correct for extinction on a star-by-star basis A$_K=(m_H-m_K-[(H-K)_\mathrm{intr}])\times1.37$, the factor is between \citet{Nishiyama_06} and the value of \citet{Fritz_11}\footnote{We note that more recent works get slightly smaller factors, see \citet{Nogueras_18}, and \citet{Nogueras_20}.}. For the intrinsic color ($(H-K)_\mathrm{intr}$) we use 0.25 as typical for bright giants. 

We exclude stars which are clearly in front of the nuclear region to make our survey more efficient. Such foreground stars are less extinguished and thus more efficiently targeted with other spectrographs. 
We use a cut in $H-K$, which varies by field from 0.3 to 0.9.  
This is an intentionally blue color cut that aims to include all nuclear disk stars.

We probably include some foreground stars this way \citep{Nogueras_19c}, but it is easier to exclude an observed target a posteriori than to
correct for missed stars. 
Using this color cut, on average we exclude 1\% per field and at most 4\%. In Schultheis et al. (submitted) we use stricter cuts in color which use spectroscopic information on star properties together with dynamic information to construct purer samples. Ultimately, the best approach will be probabilistic using all available information (i.e. colors, magnitudes, line-of-sight velocities, temperatures and metallicities).
After that step we make a selection by observed magnitude, we exclude sources with m$_K>14$, to limit the exposure time.

This magnitude selection effects only a few sources in the catalog, 
on average 0.1\% per field, at most 0.7\%. 
In practice there are probably more omitted very red sources, because we omit stars which do not have H-magnitudes from all catalogs. One reason for a missing H magnitude is that a source is too faint for detection. Similarly, a source can already be missing from the K-band catalog. To avoid a large impact due to this we shifted a few fields slightly to exclude line-of-sights towards infrared dark clouds, for example we shifted a field away from the symmetry axis to l/b -0.156/0.173\degree\, for that reason.

Since our program is designed for bad seeing, we excluded sources with a close neighbor\footnote{By mistake we cleaned the catalog which had already been cleaned from foreground sources, therefore we observed a few sources with a close by foreground sources. The number is small enough to not affect the analysis.}. 
A star is excluded when it has a close neighbor located within 2\arcsec and $<$0.5 mag fainter in K;  within 2-2.5\arcsec and $>$0.5 mag brighter; or within 2.5-3\arcsec and $>$3 mag brighter.
In total that results in the exclusion of 3\% to 18\%
(on average 9\%) of the previously defined sources, depending mainly on source density, such that more sources are affected close to the Galactic center. Overall, our selection results in between 229 and 1283 main target stars per KMOS field.
We add APOGEE sources within the fields for calibration. In total, there are seven sources for the 29 patrol fields. Only four of those sources were actually observed, because not all patrol fields were observed.

  \begin{figure}
   \centering
   \includegraphics[width=0.72\columnwidth,angle=-90]{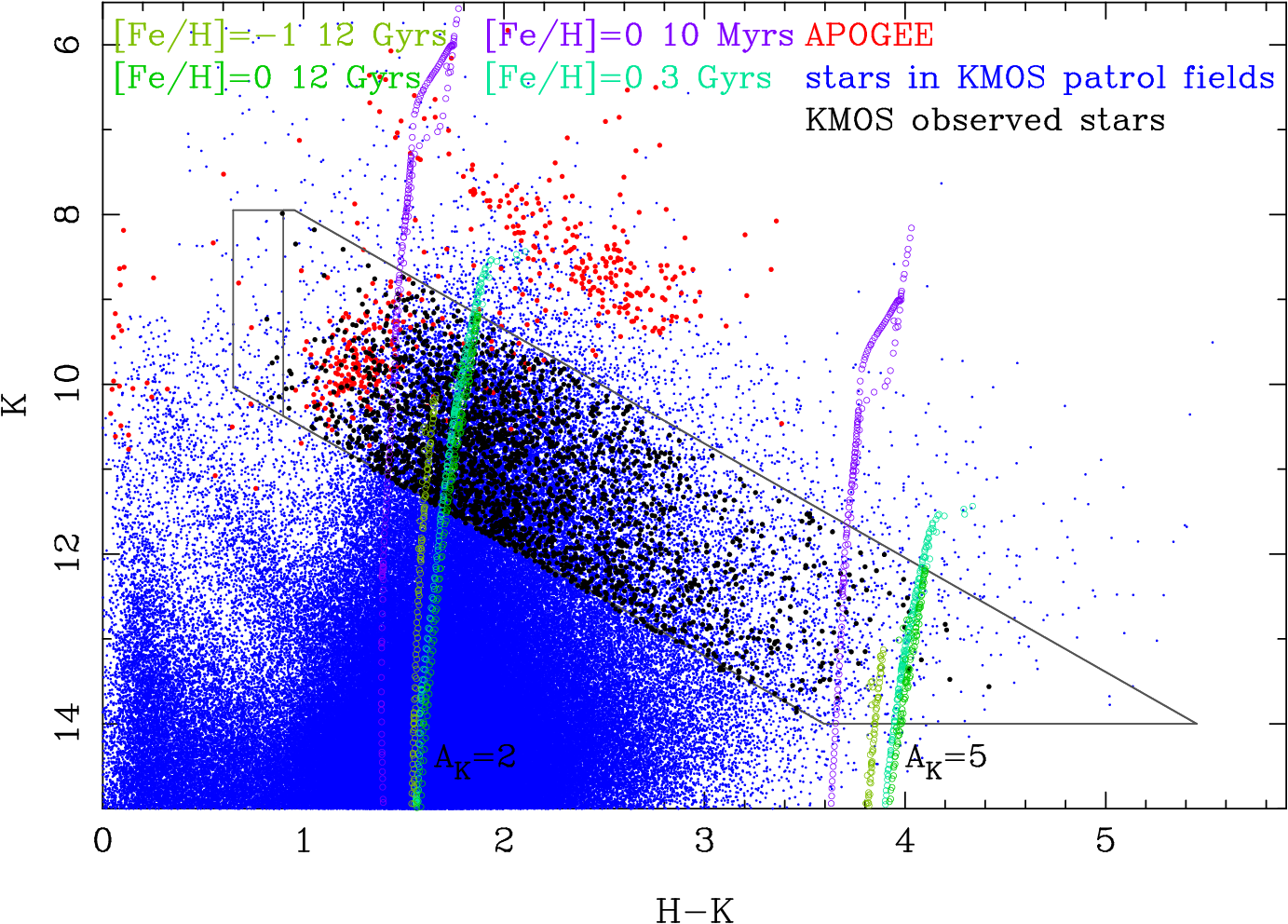}  
   \includegraphics[width=0.72\columnwidth,angle=-90]{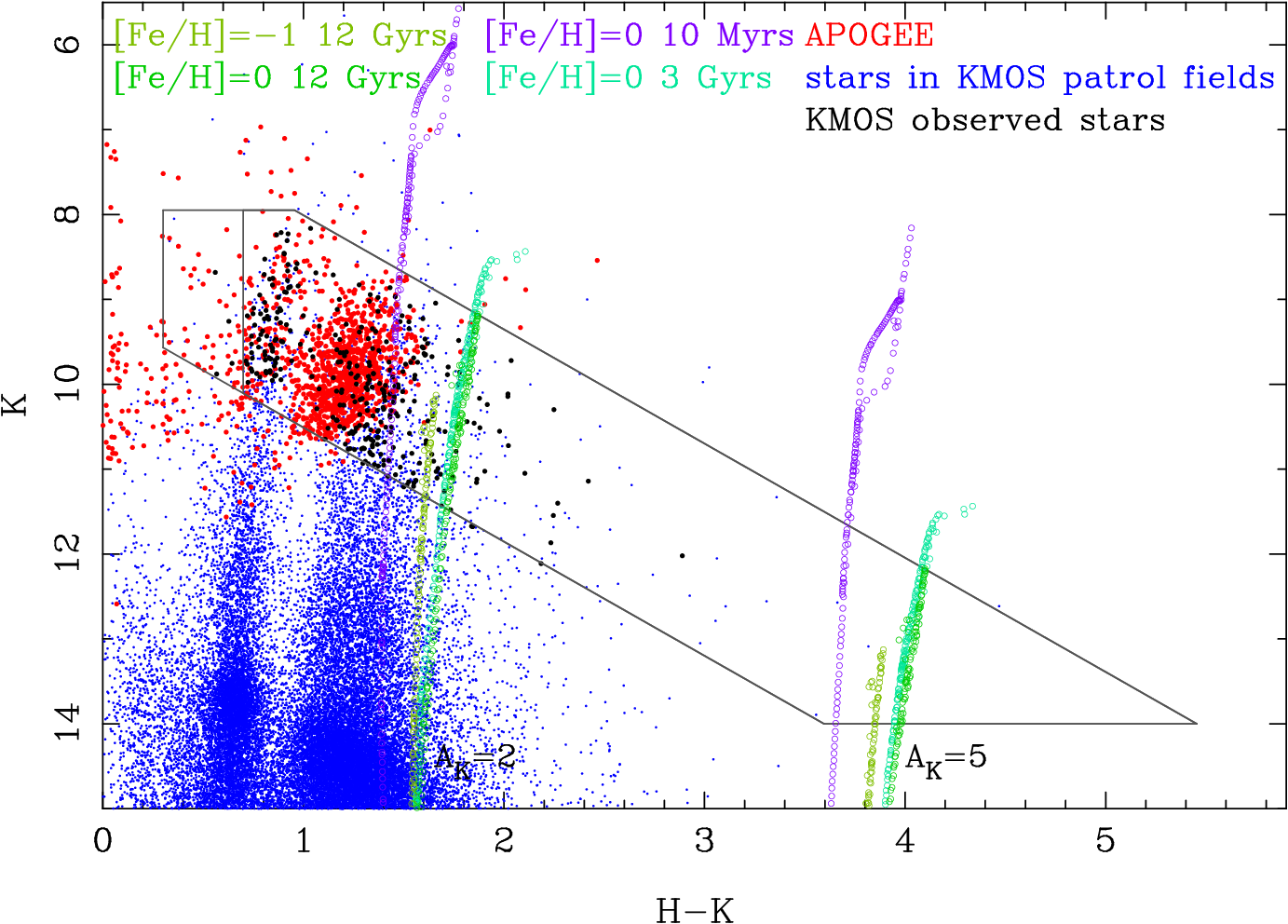}  
      \caption{Observed CMDs of nuclear disk based on SIRIUS/2MASS (top, $|l^*|<1.55\degree$ and $|b^*|<0.3\degree$) and inner bulge (bottom $|l^*|<2\degree$ and $0.3\degree<|b^*|<1\degree$. There are two sequences in the KMOS bulge data because of the variable extinction, see also \citet{Nogueras_18b}. In blue we show all stars within the KMOS fields. The KMOS and APOGEE target stars are shown in black and red respectively. The shown isochrones are from PARSEC. The gray polygons outline our CMD selection of targets. The blue H-K color limit varies slightly from field to field.
              }
         \label{fig:cmd_nuc}
   \end{figure}

The details on the observations are in Appendix~\ref{ap:field_details}. We show in Figure~\ref{fig:cmd_nuc} the CMD of our observed stars together with all stars in our field and the APOGEE targets in the inner Galaxy, all separated in nuclear disk and inner bulge\footnote{For APOGEE we use all in
 DR16 \citep{Ahumada_19} published sources, that includes besides main survey targets also telluric targets and targets of other observations. Besides the outlined spatial selection, we require K-magnitude and S/N$>14$. Below this number even line-of-sight velocities are often unreliable.}. For the star selection in nuclear disk and inner bulge we use shifted Galactocentric coordinates ($l^*$ and $b^*$) such that Sgr~A* is at 0/0.
It is clear that, towards the nuclear disk, APOGEE only targets stars brighter than the old population or stars at extinctions clearly smaller than the mean. Probably nearly all of the latter are bulge stars projected onto the nuclear disk.

\subsection{Design of the KMOS observing blocks}
\label{sect:surv_kmosobs}
 
To observe a sufficiently large number of stars per field, we planned to observe five different KMOS configurations per field with varying targets distributed over the 24 KMOS arms. Therefore, we divide the above defined catalog by K-band magnitude into five different sub catalogs. We use the K-band magnitude for reducing the dynamic range of each observing block (OB), to make it easier to avoid saturation and insufficient signal to noise ratio (S/N). In the following, one such selection is called a subset. 
In part, not all subsets were observed successfully. 

\begin{figure}
   \centering
   \includegraphics[width=0.72\columnwidth,angle=-90]{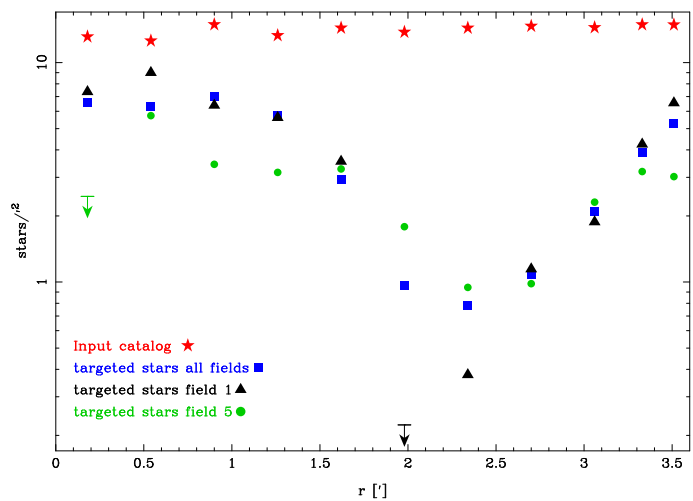}     
      \caption{Number of targets per square arcminute in a KMOS OB as a function of radial distance in arcminutes from the field center.      
      The overall target numbers are divided by 24 to show the typical target density per KMOS field. Fields 1 and 5 are  the most extreme fields in terms of target density, field 1 is at the higher density limit, field 5 is at the lower density limit and has also fewer observed stars because one subset was not successfully observed. The downward-facing arrows mark bins in which no star was observed. 
              }
         \label{fig:tar_dens}
   \end{figure}

From the sub catalogs the targets are
selected by the arm allocation tool KARMA using the Hungarian algorithm. 
The number of potential targets is usually high enough that all arms can be allocated; 
only in a few cases 1 to 3
arms are not allocated.
The algorithm results into an uneven spatial sampling, the center of a field (r$<1.4$\arcmin) is targeted most densely, then there is a minimum at about 2.3\arcmin\, from where it rises to the outer rim to nearly the same level as in the center. This effect is stronger with a larger catalog size.
We show the effect in Figure~\ref{fig:tar_dens}. 
For our science aims this uneven sample within a field is not important.

Because the stellar density is high in the Galactic center it is difficult to obtain an appropriate sky position for each individual OB, see e.g. \citet{Feldmeier_17a}, even for such short exposures as those used here. Therefore, we observed a dedicated sky field centered on the dark cloud at l/b -0.0654/0.1575$\degree$. This ensures that we encounter the minimal number of bright stars in each KMOS IFU in our sky fields. 
  Three IFUs (18, 19 and 24) could not be allocated to a star-free sky area, and therefore we do not subtract any sky cube from these IFUs. The general subtraction of a spectrum constructed from nearly object free pixels in the science cube, see Section~\ref{sect:sta_spec_extrac}, acts as a first order sky correction. We check for the effect of missing sky cube subtraction in Appendix~\ref{sect:spec_nsky}.

For efficiency we do not dither for most targets and obtain only a single object and sky exposure. Only for the lower extinction bulge fields we add some dithers since the lower exposure times make that possible. 

For the telluric data we use the standard observations provided by the observatory. These are observed with the standard 3-arm telluric routine.

\section{Data}
\label{sect:data}

Here we present the procedure of reduction from raw data to 1D spectra. We present the data reduction, measure the quality of the cubes and then extract the 1D spectra of the stars in the cubes. Details and potential problems of the spectra, OH lines and continuum ripples, are discussed in Appendix~\ref{ap:spec_details}.

\subsection{Data reduction}
\label{sect:spec_red}

We use close to standard setting for most of our reduction with mainly the software package SPARK \citep{Davies_13}. We change the wavelength sampling to 3072 pixels (from 2048) to better sample the wavelength scale. For each day, we use the matching detector calibrations. We assign by hand the matching sky to the object OBs, since the headers of our sky OBs have a science setting. We use the mode sky\_tweak in the object file reduction step to optimally subtract the sky emission lines. Since we have usually only one exposure per target, cosmics  
are a problem. We correct for them with LA-cosmic \citep{Vandokkum_01}. We use the IDL variant for spectral cubes by Richard Davies\footnote{http://www.mpe.mpg.de/~davies/downloads/lac3d.tar
} and run them on the final cubes. The same code also creates the noise cubes, used in the following. 

As our science goals require the analysis of various absorption features
a good correction of the telluric transmission is important. We use for the telluric correction, 31 standards observed in the nights of the science observations. We extract spectra and noise spectra from the telluric cubes. We use all spatial pixels since the S/N is high and it improves flux calibration. We treat each of the three spectrograph sub-system groups of IFUs (1 to 8, 9 to 16, 17 to 24) separately in the following because a different telluric spectrum is available for them. Our first sample includes all observed telluric standard stars besides the B emission line stars observed (HD 186456 and HD 171219).  
It consists mainly of early-type stars between B5V and A0V, one early white dwarf (HD2191) and 3 dwarfs between G0V and G2V. The spectrum of HD 194872, classified as G3V, looks like a colder spectral type and we thus exclude it, because such stars are difficult to model and are also too similar to the science targets. 
We first correct the telluric spectra for the spectral features of the stars. 
For this, we use the ASPCAP model spectra \citep{Garcia_16} for the observed dwarfs of the same spectral types.
 We fit the best fitting template to the observed spectra  between 2.146 and 2.187 $\mu$m with the free parameters: normalization, velocity, and Gaussian smoothing width. By using this wavelength range we are concentrating on Brackett $\gamma$, the strongest line, but we check by eye that other lines are also well fitted. Before fitting we multiply by the closest ATRAN \citep{Lord_92} model spectrum\footnote{We use spectra from https://www.gemini.edu/sciops/telescopes-and-sites/observing-condition-constraints/ir-transmission-spectra.} and eliminate any telluric affected wavelengths from the fit in order to avoid that we fit the atmospheric features and not the intrinsic spectral features.  
We require Gaussian smoothing to account for the strong rotation of early-type stars, and the pressure broadened white dwarf spectrum.
For the G dwarfs no smoothing is necessary. We flux calibrate the corrected spectra using 2MASS magnitudes, blackbodies matching the temperatures of the observed spectral types and a flux of $4.288\times10^{-10}$ W/m$^2$/$\mu$m for a magnitude of 0 at the reference wavelength of 2.157 $\mu$m. 
Between the different spectral types there is up to 5\% ratio variation over the Ks-band.

 We do not use these transmission spectra directly because they often deviate too much in the atmospheric parameters (airmass and water vapor) from the science data. We use the telluric spectra to derive how each spectral pixel varies with airmass and with integrated water vapor. For the latter we use airmass times water vapor. For that we normalize all spectra, by taking the median of not strongly varying pixels. These pixels are identified in an iterative process, which gets more exclusive with the iterations. We then fit each spectral pixel with a linear model of airmass and integrated water vapor. This model is sufficient in general. It is not perfect when the transmission becomes non linear, as it is the case for high airmass at wavelength with low transmission, like for example around 2.02 $\mu$m. However, the flux there is anyway so low that it cannot easily be used for analysis, thus the impact is minor. From the scatter of the 31 spectra compared to the fit we get the error. We find the median residual of the transmission is 1.3\%, in average it is 1.5\% and at most 7\%. 
 
 Now we identify the telluric standard closest in atmospheric conditions to the science data. For that we determine the relative strength of the typical strength in airmass and water vapor. We find that the trend in airmass is 4 times stronger. 
Therefore we weight the airmass of science and tellurics up by a factor four and then choose the best match by calculating the distance in airmass-water vapor space to all and choosing the smallest one. We then change that spectrum
 according to our linear model of airmass and water vapor, to account for the airmass and water vapor differences between telluric and science observations.

\subsection{Cube quality}
\label{sect:surv_qual}

Overall 24 of 29 fields were observed. In two fields (5 and 20) one of the five subsets was not observed.
 Two subsets in field 20  have a lower S/N, since the stars were not centered  within the IFUs during acquisition.  
Due to bad seeing in them, the stars extend into the IFUs, and thus we still could extract spectra.
 
The first 10 observed subsets (fields 14 and 20) had one target less, because arm 3
has not been active.
The spatial resolution is usually good, though it is worst in field 20, where the FWHM is about 1.3\arcsec\, in the three subsets where the star center is in the IFUs, in the others it is difficult to determine. Excluding these fields, the FWHM is always less than 1\arcsec, and can be as low as 0.33\arcsec, we note that when it is so small, pixel digitization could impact this estimate. Overall the median FWHM is 0.57\arcsec.

\subsection{Star spectra extraction}
\label{sect:sta_spec_extrac}

We extract stellar spectra using a dedicated spectral extraction routine, which subtracts a local background before extraction of the object spectrum.
This is important because the background stellar emission of similar sources can lead to an underestimation of the dispersion \citep{Luetzendorf_15}.
Also in case of several sources, a more targeted extraction is necessary in an IFU.
We collapse the (non-telluric corrected) cubes into an average image and work with this image in the way explained below.
We start with extraction of the pixels of the primary object. This is identified by the brightest pixel. 
We fit the PSF shape by a two dimensional Gaussian using a 3 pixel radius box around it. 
We use the Gaussian fit to identify the pixels with more than 50\%  of the maximum stellar flux, they  represent our usual estimate of object pixels.
With the 50\% limit, we achieved maximum S/N for most noise regimes and PSF shapes. If that methods results in 3 or less pixels
we add the next brightest pixels, until we obtain 4 pixels. Thus, we avoid an undersampled PSF, which leads to strong ripples in the continuum (see also Appendix~\ref{sect:spec_ripp}).
 If the Gaussian PSF fit failed, we use pixel count ordering. Neighboring pixels are first checked to see whether they are above the flux cut. Then all pixels are checked whether they have at least 50\% of the flux of the brightest pixel.
For the background, we simply use the faintest pixels of the collapsed cube. We use at least 25\% of all pixels. This number is increased, when the object covers many pixels to avoid that the overall S/N is limited by the background. The same background pixels are used for secondary sources. Secondary sources are local maxima compared to all neighboring pixels, which are not within a primary source or the background and have sufficient S/N. The S/N cut  
is chosen such that essentially all sources for which spectral features are detectable are included. 
However, we did not include all sources that were detectable in the continuum. At most three secondary sources were found per IFU. We then add pixels from the surrounding ring of pixels when the following conditions are fulfilled: They are not already selected for a source or background and they are brighter than 50\% of the main secondary pixel.

After that first extraction, we make three quality control checks on the primary sources to make sure that they are the targets selected from the catalog.
\begin{itemize}
  \item Firstly, we compare the spectroscopic flux, and spectroscopic color against the photometric properties from the input catalog.
  We also check the sources where the primary source has less than 50\% of the flux of all sources in the cube.
  \item Secondly, we  check the target pixel coordinates against the typical target pixel coordinates in the other IFUs in that exposure to identify offset sources.
  \item Lastly, we also check all sources whose S/N is below 10.
\end{itemize}
 
When one of these three checks has a negative outcome we look at the collapsed cube and, if we find that the current extraction is clearly suboptimal, we change object pixels by hand. For a few collapsed IFU cubes we also identified additional secondary sources during that process.
In case of sources with more than one exposure we also ensured that the order of sources is always the same. 
For some sources we saw in the collapsed cubes that the background include bad pixels.
All these changes affect only 37 sources (i.e. 1.2\% of all targets).

\section{Spectroscopic analysis}
\label{sec:spec_ana}

Here we describe how we measure basic properties from the spectra. We measure the line-of-sight velocity and spectral indices for H$_2$O, Brackett $\gamma$, Na I, Ca I and CO 2-0.

\subsection{Velocity measurements}
\label{sec:vel_meas}

As a first step we obtain radial velocities because they are needed for the subsequent analysis of the spectra.
Most stars show CO band heads, and we use cross-correlation at 2.18-2.425 $\mu$m to measure line-of-sight velocities, because the band heads are too complex for other ways of analysis. 
We normalize our spectra to 1 by fitting linear function to the spectra before the CO band heads (2.08 to 2.29 $\mu$m). 
As templates we use GNIRS spectra from Gemini\footnote{http://www.gemini.edu/sciops/instruments/nearir-resources/spectral-templates/library-v20} which cover at least from 2.18 to 2.425 $\mu$m. We select mostly very late giants and some earlier giants between F7 and M0. For most stars the best correlation is achieved with the K7III star, HD63425B. 
We correlated all KMOS spectra with all templates. Usually very similar velocities are achieved for all the different templates. Then we correlated the template with the largest correlation coefficient in wavelength sections with the spectra. In most cases, we select six sections, one before the band heads and then one for each band head. In all but 43 cases the velocities in the different sections agree very well. The others we inspect closer, a few are clearly late-type but have positive spikes (and a few negative ones) likely due to sky subtraction problems. We correct them like bad pixels. 
Further, we correlated the problematic spectra in three sections, one before and two in the band heads. After that process all primary spectra except for 19 have a velocity consistently determined over the full wavelength range and in sections, with at most 30 km/s difference, usually much less. 
They also show a small scatter (at most 45 km/s) between the different sections and have also agreeing velocities for the different templates. We inspected the more diverging cases by eye. They are all late-type stars. 
Thus, velocities with large errors can also be trusted, since there is the right signal in the spectra. 
   
In total we have 3405 CO based velocities of primary spectra. Not all are from different stars, since some stars have several exposures.
For the velocity errors ($\sigma_\mathrm{l.o.s}$) we use primarily the error given by the correlation. It is usually consistent with the scatter from the velocities determined in sections although the latter should be larger when the error is dominated by S/N. That shows that we do not underestimate errors, and that for most stars the S/N does not matter for the error, but systematics like template target mismatch are the dominating contributions. 
For stars with S/N$<25$, the formal correlation errors are larger. We correct for that to obtain the velocity error over the full range by $\sigma_\mathrm{l.o.s\,segm}=\sqrt{\sigma_\mathrm{segm}^2-(c/[S/N])^2}$. 
Here $\sigma_\mathrm{segm}$ is the scatter over the velocities in the different wavelength ranges. $c$ is determined such that  $\sigma_\mathrm{l.o.s\,segm}$ is in the median the same as the error of the correlation over the full wavelength range. 
We compare the errors obtained this way with the  errors obtained by correlation. 
We used the errors obtained in this way for stars which have a large $\sigma_\mathrm{l.o.s\,segm} $ (less than 5\% likelihood by chance). That is the case for 248 spectra. We use the same method for secondary source spectra.

Stars in some bulge fields have several exposures. We combine their measurements in the following way. We only use exposures which have a velocity measurement.
Of them we combine the different velocities error weighted. In the median the error derived from the error weighted scatter over the different exposures is consistent with the error which follows from the error weighted combination of the individual uncertainties.
We check for each star whether the $\chi^2$ is within the 95\% quantile for the available number of spectra. That is the case for 201 stars, and not for 44 stars. For them we upscale the error by $\sqrt{\chi^2/\mathrm{d.o.f.}}$. The upscale factor is at most 3.6 and the final uncertainty at most 17 km/s which shows that our statistical errors are well estimated and small. Their S/N is calculated as $\mathrm{S/N}_\mathrm{exp}\times\sqrt{\mathrm{EXP}}$. 

\begin{figure}
   \centering
   \includegraphics[width=0.72\columnwidth,angle=-90]{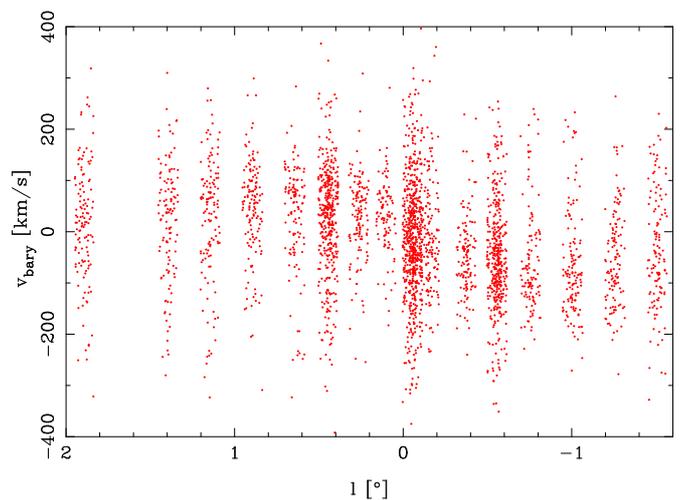}   
      \caption{Barycentric velocities of all stars with measurements as function of Galactic longitude.
 }
         \label{fig:vel_l}
   \end{figure}

Overall we have CO-based velocities for 2790 primary sources and for 241 secondary sources. 
We obtain velocities for the early-type stars by fitting their lines, see Patrick et al. in preparation.
For the primary sources there are six spectra without velocity, five have low S/N
below 12, one has higher S/N and is feature free. We discuss them more in Patrick et al. in preparation, together with the other young stars.
Of the stars with CO lines the median velocity error is 2.5 km/s, the 5\% with the largest error have errors larger than 6 km/s, the 1\% with the largest error have errors larger than 11 km/s, and the largest error is 44 km/s. 
Secondary sources perform somewhat worse, as expected due to the lower S/N:
Of them 40 spectra have no velocity, out of which 33 have low S/N
below 24 and seven have higher S/N, and are some kind of early-type star. Of the secondary stars with CO lines the median velocity error is 4.5 km/s,  5\% with the largest error have errors larger than 20 km/s, and the largest error is 54 km/s.
We test the velocity calibration and errors (see Appendix~\ref{sec:vel_tests}) and make small corrections because of it. An overall shift of 4 km/s for the velocities and we set a lower limit of 4.2 km/s for the error.
We show the velocities as function Galactic longitude in Figure~\ref{fig:vel_l}.

\subsection{Spectroscopic indices}
\label{sec:vspec_indices}

We measure for all stars spectroscopic indices for Brackett $\gamma$, Na I, Ca I, CO 2-0 and H$_2$O. For obtaining the wavelengths in the reference frame of each star we use the velocity previously determined without barycentric corrections. For stars without a velocity measurement we assume a velocity of 0. The wavelength ranges for the indices are summarized in Table~\ref{tab:spec_ind}.

For the CO index we use the index definition of \citet{Frogel_01}\footnote{We change the sign definition for this as for the other indices, such that an absorption line has negative value, and an emission line a positive value.}. For the continuum reference we fit the four continuum ranges by a linear function in which we give all ranges the same weight. 
We then integrated over the CO band range without interpolation and convert the result to the equivalent width in \AA. 
The error is estimated from repeated observations of the same star. It is 57/(S/N). This could be an underestimation when systematics like extinction or velocity play a role. The index of  \citet{Frogel_01} is not particularly sensitive to those \citep{Pfuhl_11}, but a small contribution is possible.

 \begin{table}
      \small
      \caption[]{Vacuum wavelength ranges used for the spectroscopic indices. For CO 2-0, Na I and Ca I we use the definitions of \citet{Frogel_01}, while the Brackett $\gamma$ and H$_2$O index range are defined by us. The details about the indices are explained in the text.}
         \label{tab:spec_ind}
      $$
         \begin{array}{p{0.41\linewidth}l}
         \hline
\hline
  \rm{Purpose~of~wavelength~range} &  \rm{wavelength~range~} [\mu m]   \\
           \hline  
\hline
\hline
\rm{CO~continuum} & 2.2300-2.2370 \\ 
\rm{CO~continuum} & 2.2420-2.2580 \\ 
\rm{CO~continuum} & 2.2680-2.2790 \\ 
\rm{CO~continuum} & 2.2840-2.2910 \\ 
\rm{CO~feature}  & 2.2910-2.3020 \\ 
\hline
$\rm{Br~}\gamma\rm{~continuum}$ & 2.13112-2.15112\\ 
$\rm{Br~}\gamma$  & 2.16412-2.16812\\ 
$\rm{Br~}\gamma\rm{~continuum}$ & 2.18112-2.20112\\ 
\hline
\rm{Na continuum} & 2.1910-2.1966\\ 
\rm{Na~feature}  & 2.2040-2.2107\\ 
\rm{Na continuum} & 2.2125-2.2170\\ 
\hline
\rm{Ca continuum} & 2.2450-2.2560\\ 
\rm{Ca~feature}  & 2.2577-2.2692\\ 
\rm{Ca continuum} & 2.2700-2.2720\\    
\hline
\rm{H}$_2$\rm{O~feature}  & 1.9850-1.9990\\ 
\rm{H}$_2$\rm{O~continuum} & 2.1800-2.2040\\    
\rm{H}$_2$\rm{O~continuum} & 2.2107-2.2577\\ 
\rm{H}$_2$\rm{O~continuum} & 2.2692-2.2910\\ 
    \noalign{\smallskip}
           \hline
       \end{array}
       $$
       \normalsize
  \end{table}

\begin{figure}
   \centering
   \includegraphics[width=0.72\columnwidth,angle=-90]{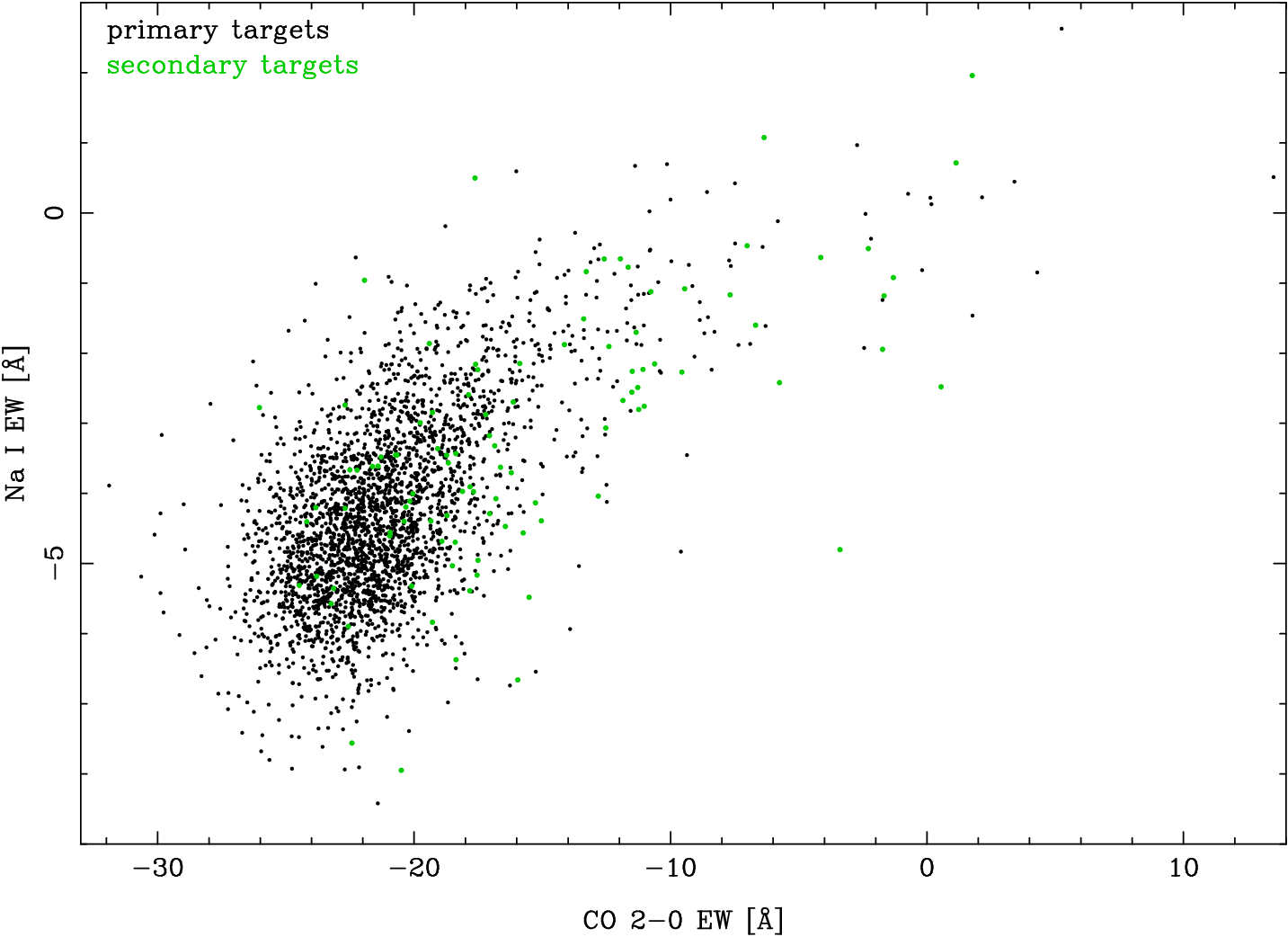}     
    \includegraphics[width=0.72\columnwidth,angle=-90]{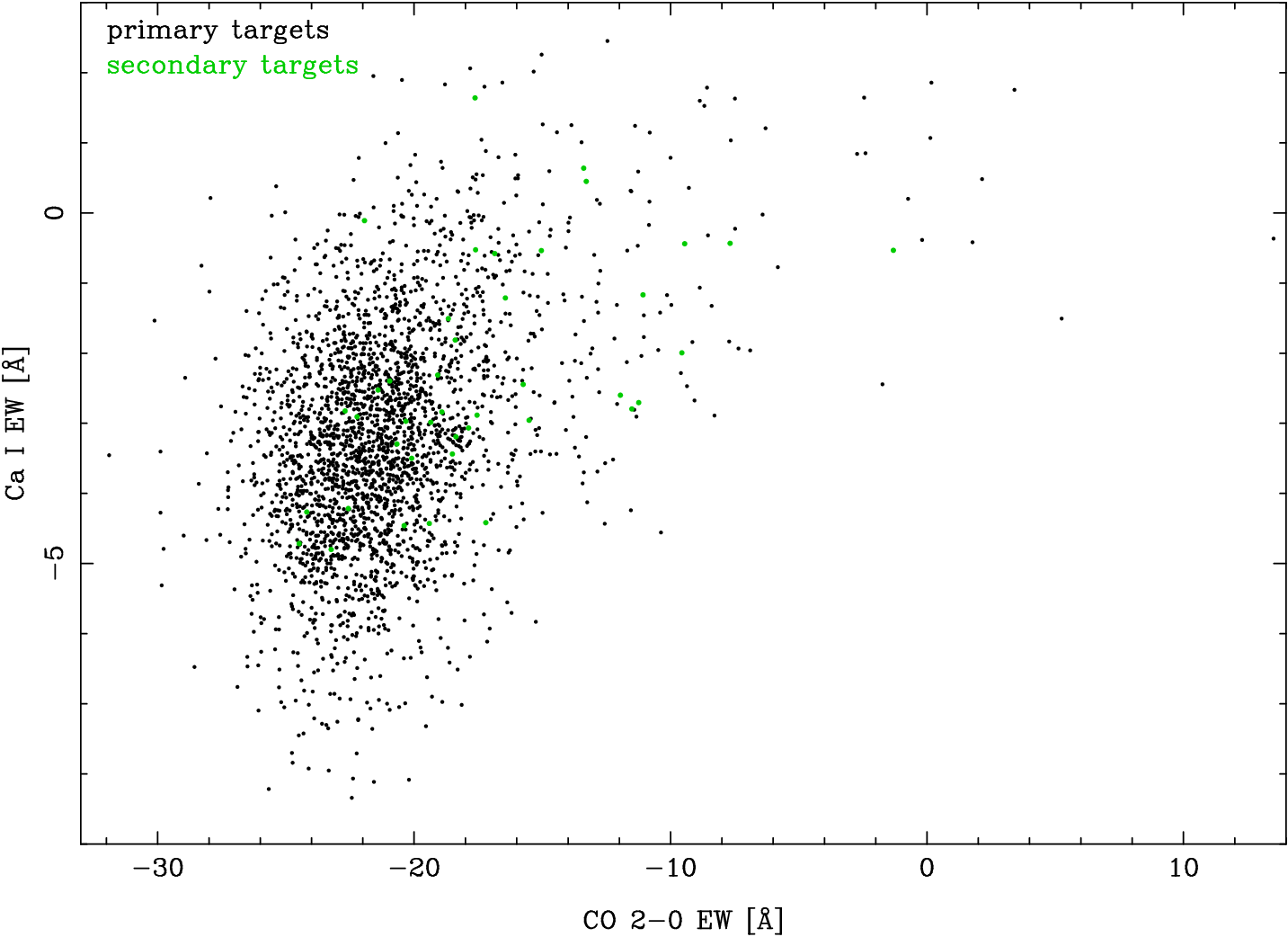}   
      \caption{Na (top) and Ca (bottom) as function of CO index. For Na we show all stars with a S/N$>20$, the value where the intrinsic scatter of CO and Na is of the size of the measurement error. For Ca the S/N is lower and thus we plot only stars with a S/N$>40$.
              }
         \label{fig:naca_indices}
   \end{figure}

For Brackett $\gamma$ we construct our own index, see Table~\ref{tab:spec_ind}. 
For this index we obtain from repeated observations of the same star an error of 23/(S/N).
We also calculate the Na I and Ca I indices of \citet{Frogel_01} for our stars. From the repeat observations we estimate an error of 27/(S/N) for the Na index and of 58/(S/N) for the Ca index. We show these indices as function of CO, which is a good approximation for effective temperature, in Figure~\ref{fig:naca_indices}.

\begin{figure}
   \centering
   \includegraphics[width=0.72\columnwidth,angle=-90]{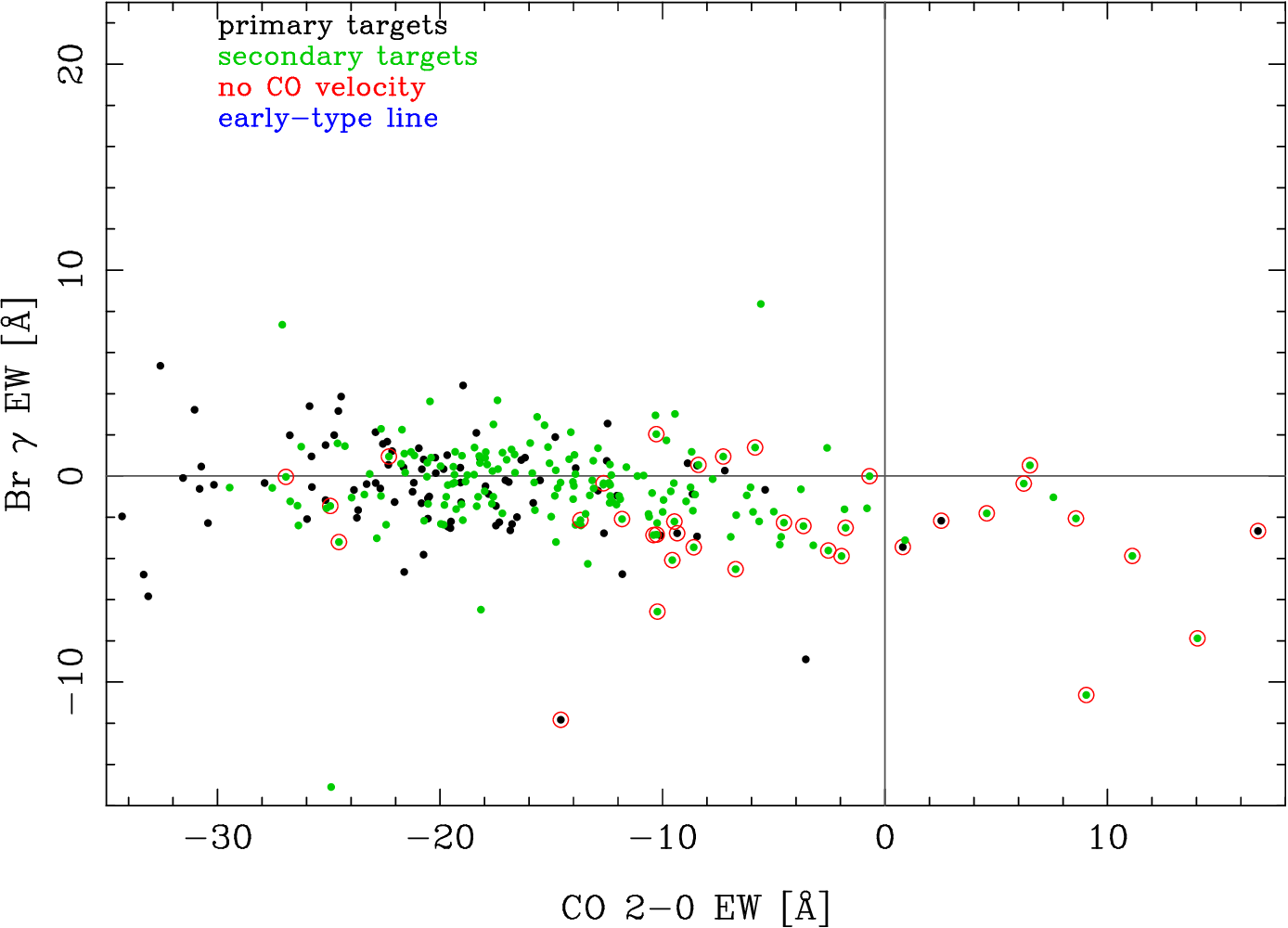}     
    \includegraphics[width=0.72\columnwidth,angle=-90]{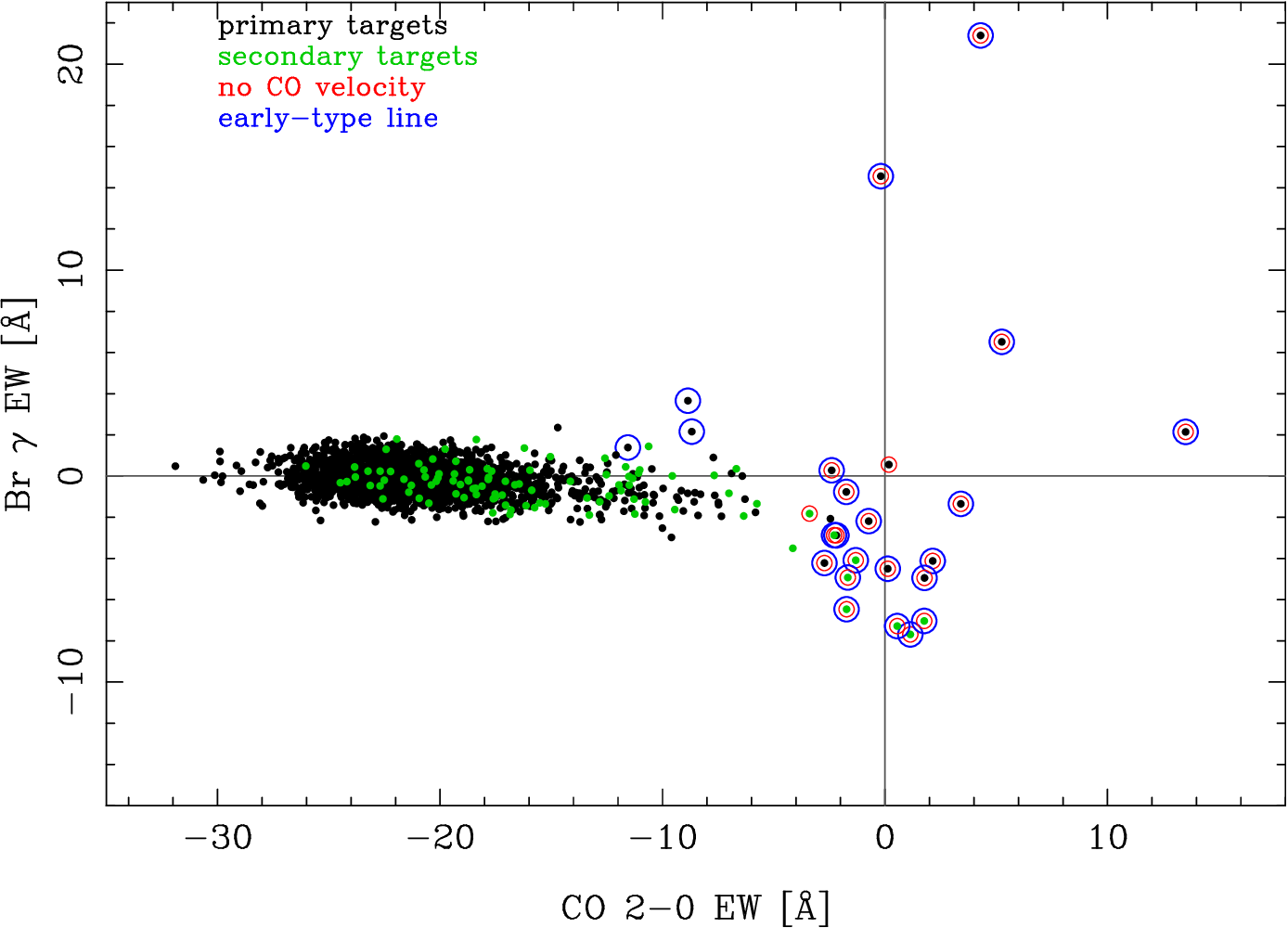}   
      \caption{CO and Brackett $\gamma$ equivalent widths for all stars. Stars with early-type lines are confirmed with visual inspection of the spectra. In the top panel we show the stars with a S/N$<20$, in the bottom one the stars with a S/N$>20$. This S/N is the approximately border of the nearly complete regime of spectral classification. The lines divide absorption (negative) and emission.
              }
         \label{fig:indices}
   \end{figure}

We show in Figure~\ref{fig:indices} the CO and Brackett $\gamma$ indices. Even when all (including low S/N) stars are plotted, the expected structure is visible. Most stars have strong CO absorption and essentially no Brackett $\gamma$ absorption or emission, thus EW$_{\mathrm{Br~}\gamma}\approx$ 0. These stars are too cold for hydrogen lines.
 Moving to weaker CO there is then weak Brackett $\gamma$ absorption. When the index is larger than about -4 \AA~most high S/N stars have then no CO based velocity. These are the young stars, see Patrick et al. in prep. 
 Around EW$_\mathrm{CO}\approx-5$ \AA~several stars have weak CO absorption and narrow Brackett $\gamma$ absorption, that is expected for warm stars. They are not early-type stars \citep{Habibi_19}. 
Finally, we looked at stars with an CO EW index more negative than -4 \AA, to find stars unusual in Brackett $\gamma$ which have CO. We only look at stars with a S/N$>20$. We determine the median in bins and fit a quadratic function to it. We then look at stars which deviate by more than 2 \AA~from the track. For some of them the signal is spurious. 
However, we find 3 genuinely unusual stars.

\begin{figure}
   \centering
   \includegraphics[width=0.72\columnwidth,angle=-90]{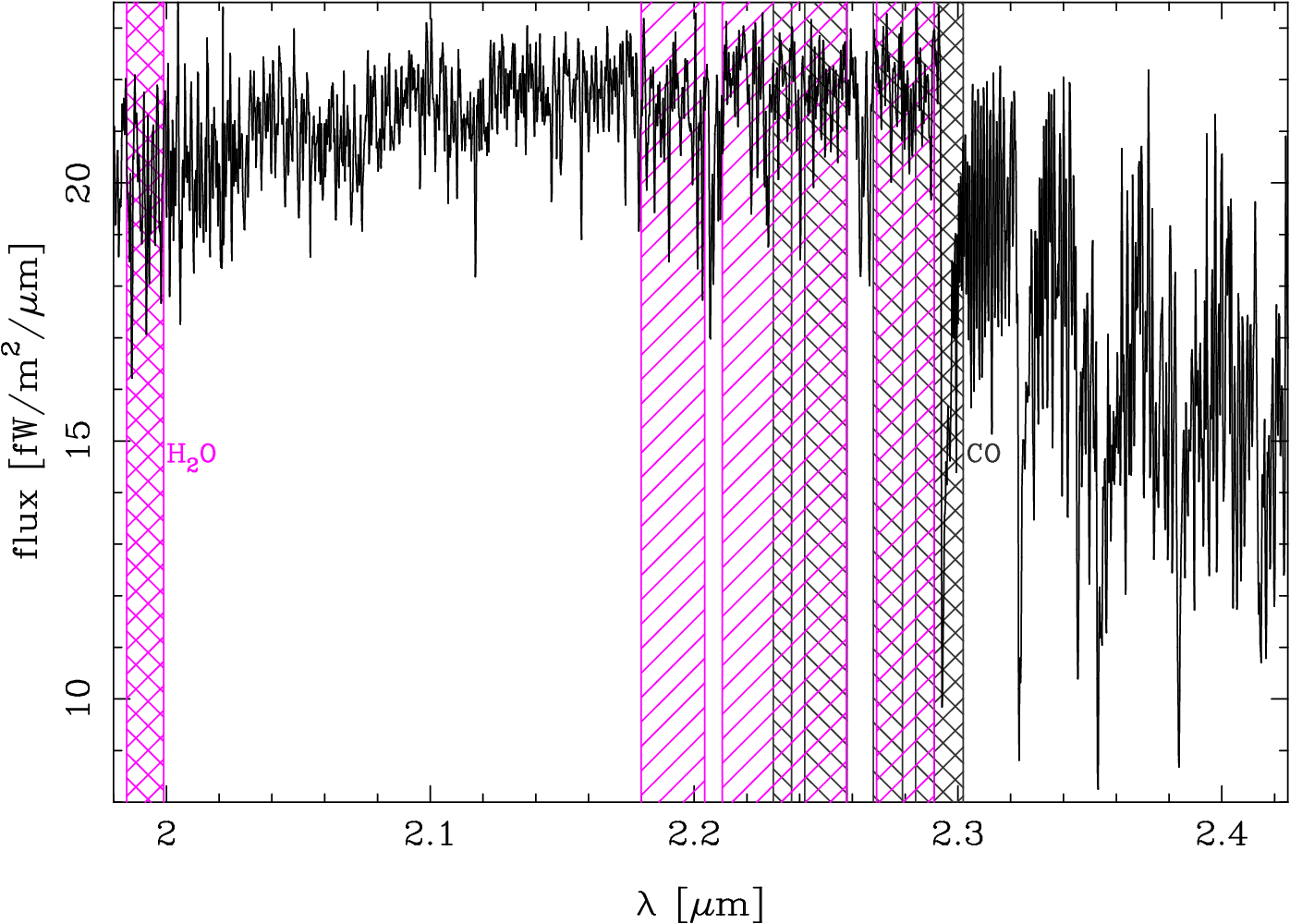}  
      \includegraphics[width=0.72\columnwidth,angle=-90]{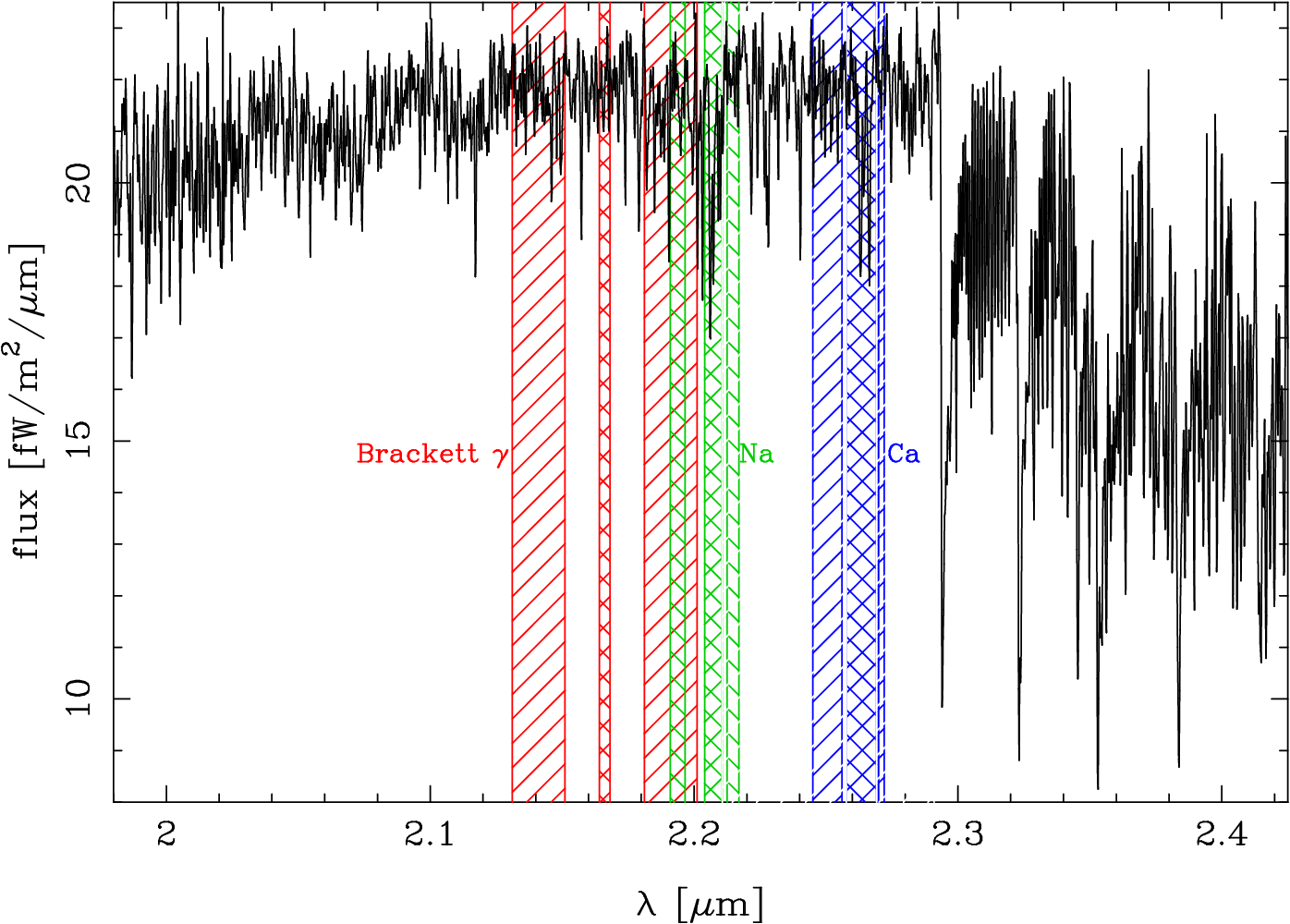}   
      \caption{Indices windows together with a typical spectrum. The continuum windows are shown hatched, the feature windows cross hatched. The spectrum is typical for a primary target spectrum in S/N and CO and Na index values. In the top panel we show the molecular indices, in the bottom panel the atomic indices. 
              }
         \label{fig:ex_spectrum}
   \end{figure}

 We check these and all stars for being AGB stars. Therefore, we look for broad H$_2$O features as typical for AGB stars, see e.g. \citet{Lancon_00}. Because our spectra cover only the K-band we cannot construct the usual indices which require H and K coverage like in e.g. \citet{Blum_03}. We construct our own index in the following way: we fit the continuum between 2.18 and 2.291 $\mu$m (excluding Na and Ca features) with a linear function of $\log{\lambda}$. This is the wavelength range where neither H$_2$O nor CO has features. The extrapolation of the fit gives the expected continuum. Then we measure the H$_2$O in a full width window of 0.014 $\mu$m around 1.992 $\mu$m. This is the bluest wavelength range where the atmospheric transmission is still acceptable. This and all other index windows are shown in Figure~\ref{fig:ex_spectrum}.

 The ratio of this flux and the extrapolated continuum flux gives the H$_2$O index. We check the formal error by analyzing the stars with multiple exposures. We find that the error is underestimated for higher S/N. The main reason is not clear but effects like correlated errors and imperfect telluric correction could be responsible. We add to the formal errors 0.048 in quadrature to have realistic errors over the full S/N range.

\begin{figure}
   \centering
   \includegraphics[width=0.72\columnwidth,angle=-90]{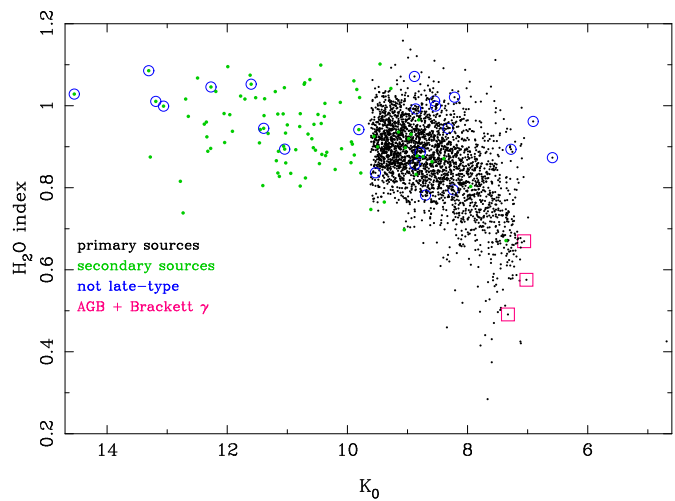}    
      \caption{H$_2$O index as function of extinction corrected magnitude K$_0$ for stars with S/N$>20$. Unusual sources are indicated with open symbols,  blue circles (early-type or young stellar objects) or boxes (AGB with Brackett $\gamma$). Sources with H$_2$O index below 0.7 are probably AGB stars. 
      }
         \label{fig:agb_co}
   \end{figure}

 We check the H$_2$O-index for dependency on extinction, by determining the median index in extinction bins. We exclude stars with K$_0$<7.5 because there are many AGB stars and thus there is a stronger variation. There is some trend over the full range, but the impact over the central 80\% extinction range 
 (A$_K$ between 1.46 and 3.49) is with 0.044 smaller than the error floor, it steepens somewhat at the blue and red end. The H$_2$O index is smaller for more extincted stars. That is expected when the stars are not all at the same distance since more distant stars suffer higher extinction and are intrinsically brighter. Brighter stars have a larger H$_2$O index in our sample in the median. Whether this can explain the full trend is unclear. It is possible that we do not fully correct for extinction effects in our measurement. It is also possible that our assumption of constant $(H-K)_0=$0.25 matters.
 Still, the index is sufficiently well determined so that plausible trends are visible, see Figure~\ref{fig:agb_co}. Most stars including the early-type stars have an index between 0.8 and 1.1.  153 stars have index below 0.7 and are strong candidates for AGB stars. They are distributed over a relatively large range in CO index between -8 and -32 EW [\AA]. The three late-type stars with Brackett $\gamma$ emission are among them and belong to the AGB candidates with the weakest CO features. A similar trend is visible in the figure S13 of \citet{Lancon_00} which is for the same star in different phases. These stars belong also to the brightest in our sample, but the luminosity could be slightly overestimated since we do not consider AGB star effects for the effective temperature and intrinsic color estimation.

 \section{Physical properties}
\label{sec:phys_properties}

   To obtain physical properties of the late-type stars (all stars with CO absorption and radial velocity measurements) we use the indices.
   We use calibration stars with known metallicity to derive an index-metallicity relation. For temperatures, we combine CO index-temperature relations from the literature.
   
    \subsection{Metallicity calibration}
\label{sec:met_cal}

Firstly, to estimate the metallicity, we test the index calibrations of \citet{Frogel_01}. \citet{Frogel_01} collected a sample of calibrators together with their  Na, Ca, and CO equivalent widths and fit the full sample with linear and quadratic functions.
We find that both their options lead to unrealistic low metallicities for stars with strong lines. The quadratic function even leads to decreasing metallicity for stronger Na at large CO EW depth. That is not surprising because the calibration of \citet{Frogel_01} is based on globular clusters and thus is not well suited for the expected metallicity range of our sample. We therefore derive our own calibration. We use Na and CO EW widths because they have the highest S/N. \citet{Frogel_01} found that Ca contributes only weakly to the metallicity estimate. We find the same. 

\subsubsection{Metallicity index construction}
\label{sec:met_cal_const}

For the derivation of our own calibration we use the following spectra. 
\begin{itemize}
  \item Firstly, 18 of our KMOS stars were also observed by APOGEE and have their metallicity determined in DR16 \citep{Ahumada_19}. 
  \item Secondly, we use the X-shooter spectral library (XSL), precisely the spectra published in \citet{Gonneau_20}. We select spectra with log(g)$<2.5$ according to \citet{Arentsen_19}. Around the order transition at 2.275 $\mu$m the XSL spectra show some spikes, we corrected for the worst by assigning them bad pixels. Also we calculate an alternative CO index which uses two continuum points redward of the problematic region. The two CO indices deviate from each other by 1.0 \AA~EW standard deviation with a median bias of 0.5 \AA~EW whereby the \citet{Frogel_01} index is slightly smaller. We use the average of both in the following. 
  A few stars have several spectra in the XSL, we compared the values obtained for them, the differences are very small, and thus they measure the same quantities. In such cases we use the median of the index values in the following way. We exclude stars with a positive (CO emission) CO index, because we are only analyzing stars with CO absorption here. Thirdly, we use SINFONI spectra of the Galactic center stars from \citet{Thorsbro_20}. For GC25 we do not have a spectrum and we exclude GC16867 because of too small S/N. A few stars have two spectra, with very similar EW values. Therefore we use their average. 
  \item Finally, we also use NGC6583-46 for which we obtained a spectrum with FIRE\footnote{\url{http://web.mit.edu/rsimcoe/www/FIRE/observers.htm}}.
\end{itemize}

For stellar properties we use  APOGEE DR16 properties when possible. We use [Fe/H] when available and use [M/H], as provided by \citet{Ahumada_19}, for three KMOS stars for which [Fe/H] is not available.  From the stars with both, we determined that [M/H] is in average 0.012 dex larger, we subtract this number from three metal rich stars which do not have [Fe/H]. APOGEE data are available for the 18 KMOS stars and for 21 stars from XSL. 
For these stars we determine that the average difference in metallicity between APOGEE and \citet{Arentsen_19} is 0.088 dex, whereas the metallicity of APOGEE is larger.
Therefore, we add 0.088 dex to all metallicities from \citet{Arentsen_19}. For the Galactic Center stars we use the [Fe/H] from \citet{Thorsbro_20}. 
We use for NGC6583-46 the high resolution spectroscopy results of \citet{Magrini_10} for this cluster. 
In general these different data sets agree about as well as expected from their errors. The exception is that cold stars from \citet{Arentsen_19} do not follow the same pattern in CO-Na space, it is more random. \citet{Arentsen_19} mention in Sect.~5.4 that they do not trust their results below 3800 K, therefore we exclude those stars. 

From all samples we exclude stars brighter than M$_K=-7.5$. We choose this limit since the nuclear disk stars sample excludes brighter stars and because beyond that limit most stars are not red giants but AGB stars and supergiants, which often have different index strengths for the same temperature. We use Gaia DR2 parallaxes \citep{Brown_18} for the local stars, and a Galactic Center distance of 8.18 kpc \citep{Abuter_19,Bland_16} for the bulge and nuclear stars. Similarly, we use distances from \citet{Harris_96} for the globular cluster stars and a distance of 50.1 and 62.8 kpc for the large and small Magellanic Cloud member stars, respectively \citep{Fritz_20}. For NGC 2324 we use a distance of 3.8 kpc \citep{Piatti_04} and for NGC 2682 we use a distance of 0.88 kpc \citep{Babusiaux_18}.
For magnitudes we use for most, 2MASS. Besides we use the previously mentioned SIRIUS magnitudes for the KMOS stars and magnitudes from \citet{Fritz_16} for the stars from \citet{Thorsbro_20}. We correct the latter stars like the nuclear stars for extinction. The others we do not correct for extinction, the extinction is usually small and it does not matter much when we include a few slightly brighter stars because not all slightly brighter stars have different indices. We are excluding stars with H$_2O\mathrm{-index}<0.7$. That excludes most AGB stars.

  \begin{figure}
   \centering
    \includegraphics[width=0.72\columnwidth,angle=-90]{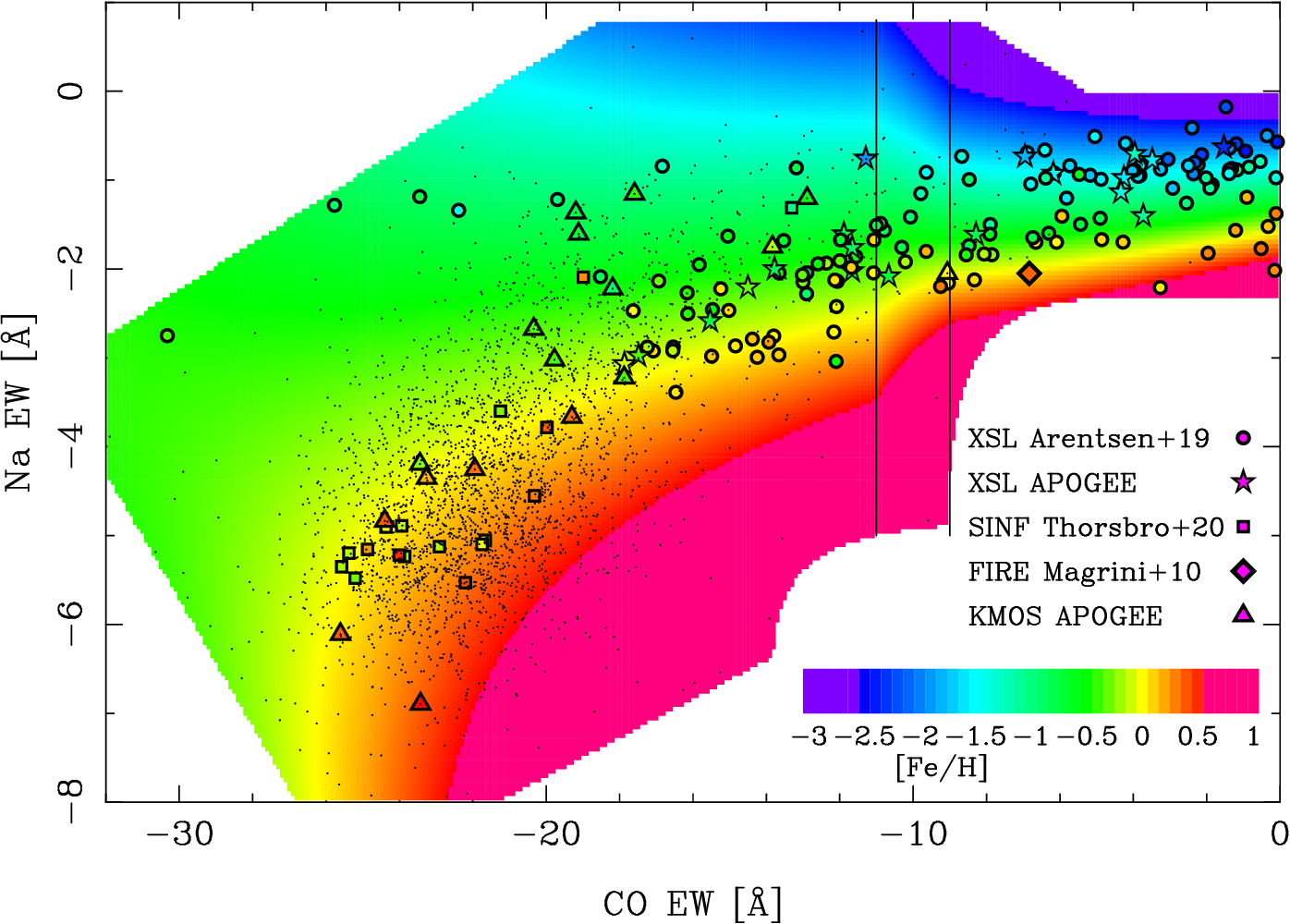}  
   \includegraphics[width=0.72\columnwidth,angle=-90]{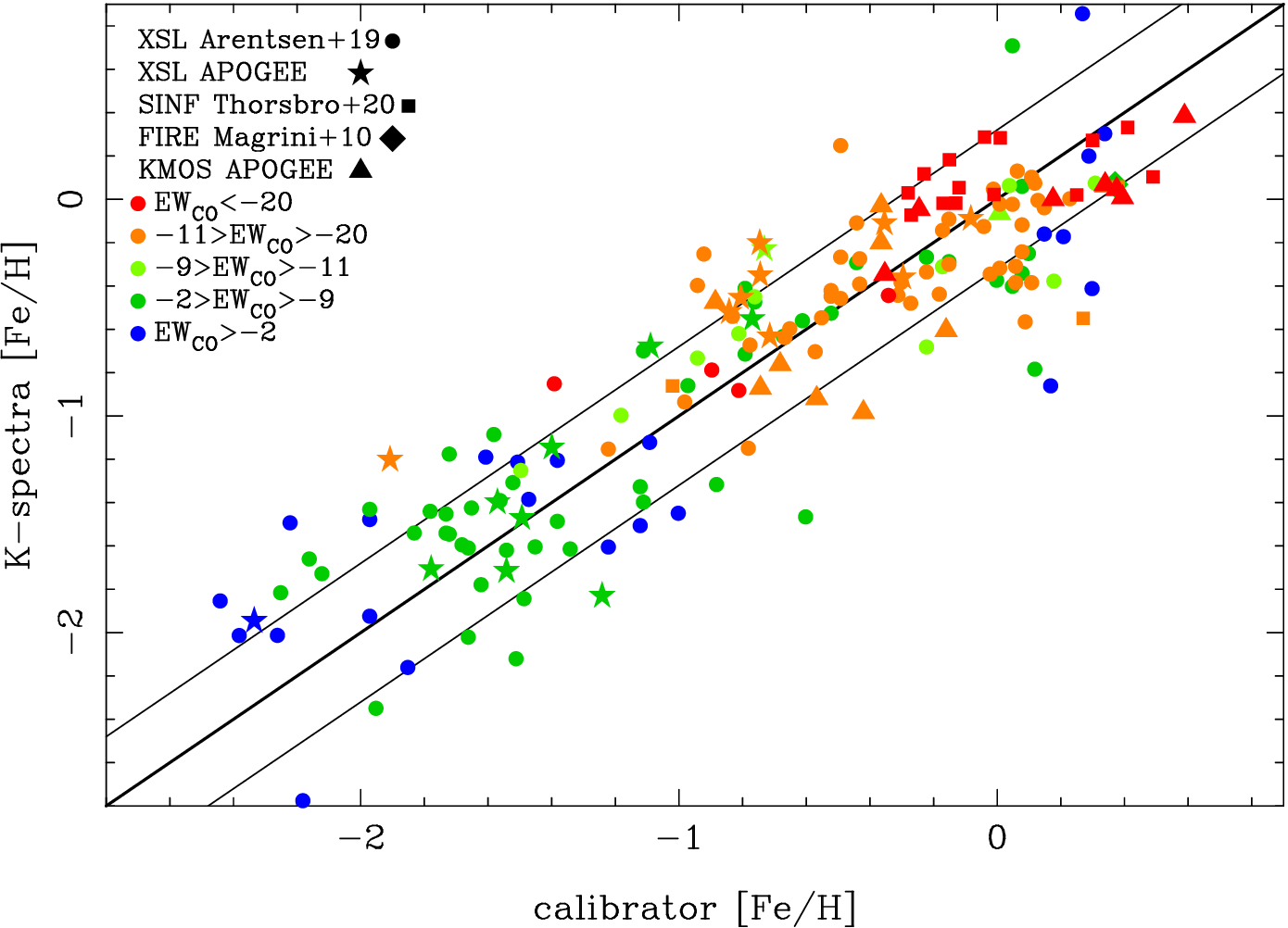}    
      \caption{Metallicity derivation. Top: Calibration of metallicity based on CO and Na EW. The large symbols are the 183 used calibration spectra, the symbol shape shows the source of spectra and metallicity \citep{Magrini_10,Thorsbro_20,Ahumada_19,Arentsen_19}, the value is indicated by the color within, SINF indicates SINFONI. The background shows the derived relation, the color range saturates below [Fe/H]$<-2.5$ and above [Fe/H]$>0.6$ (the range of the calibrators.), where the derived values are based on more uncertain extrapolation. The two vertical lines separate the application of the high and low CO solutions, in the overlap region between we use both to derive metallicities. The small dots show all our KMOS targets with S/N$>30$ and a CO based velocity. The background shows the derived value only for areas where either calibrators or the nuclear stars have coverage.  Bottom: Comparison between literature (x-axis) and our metallicities (y-axis). The thick black line is the identity line, the thin lines bracket the typical uncertainty. Stars are coded by CO depth, the hottest ones have the smallest depth. 
      }
         \label{fig:met_cals}      
   \end{figure}

All this together results in a sample of 187 stars, which are listed in Appendix~\ref{ap:met_cal}.
We tried to fit the full sample at once but found that such a high order polynomial is needed, that extrapolation beyond the data range is implausible and also within the data the metallicity does not always increase with feature strength in contrast to the expectations. Therefore we divide the data set into parts by CO strengths. Firstly, there is a shallow CO depth range, the part where the maximum possible Na strength does not increase much. That is the case for EW$_\mathrm{CO}>-11.5$ \AA, see top of  Figure~\ref{fig:met_cals}. Secondly, this is a region where both Na and CO EWs clearly vary, we choose a range down to CO EW of -8.5 \AA, to make sure both use partly the same data. That avoids a big jump in the overlap region. When applied to spectra we linearly change the weight of the two for EW between -11 and -9, to ensure a smooth overlap.
From the sample we excluded four outliers, two of them are of very low CO depth, where we do not derive metallicities from targets.

We derive the following relation. For EW$_\mathrm{CO}>-8.5$ we obtain:
\begin{equation}
\begin{split}
\mathrm{[Fe/H]}=-3.14-0.0106\,\mathrm{EW}_\mathrm{CO}-1.98\,\mathrm{EW}_\mathrm{Na} +0.00763\,\mathrm{EW}_\mathrm{CO}^2\\
-0.0929\,\mathrm{EW}_\mathrm{CO}\,\mathrm{EW}_\mathrm{Na}+0.00646\,\mathrm{EW}_\mathrm{Na}^2
\end{split}
\label{eq:met_lowco}
\end{equation}
wherein all EW are in \AA. For EW$_\mathrm{CO}<-11.5$ we obtain:
\begin{equation}
\begin{split}
\mathrm{[Fe/H]}=-1.65+0.0317\,\mathrm{EW}_\mathrm{CO}-1.07\,\mathrm{EW}_\mathrm{Na}
+0.00195\,\mathrm{EW}_\mathrm{CO}^2\\
-0.0288\,\mathrm{EW}_\mathrm{CO}\,\mathrm{EW}_\mathrm{Na}-0.0211\,\mathrm{EW}_\mathrm{Na}^2
\end{split}
\label{eq:met_highco}
\end{equation}

In the bottom panel of Figure~\ref{fig:met_cals} we show the comparison between the metallicities of our reference sample and our derived metallicities using equations~\ref{eq:met_lowco} and \ref{eq:met_highco}. 
We notice that the scatter is larger for the warmest stars with the weakest CO absorption  which is expected as for those stars the used features are weak.
The standard deviation between the input and the derived [Fe/H] is 0.32, which varies little over almost the entire range of CO depth. The scatter is 0.5 for $EW_\mathrm{CO}>-2$. It does not impact our conclusions because none of our targets stars has such a small depth; the smallest depth of a late-type star in our science sample is -2.5 when ignoring low S/N sources.

We see in addition more deviations from the identity line for metal-rich stars ($\rm [Fe/H] > 0$). This is due to the fact that cool metal-rich giants  ($\rm T_{eff} < 4000$\,K) suffer from substantial molecular lines in their spectra which results in consequent line blending and blanketing and affects the stellar parameters. The increased scatter could therefore be also due to the larger errors in the metallicities of the reference sample we are using. 72\% of our sample are cool stars with $\rm EW < -20$ which lie mostly inside the 1$\,\sigma$ scatter.
In the top panel of Figure~\ref{fig:met_cals}, it is visible that the metallicity follows the expectations: the metallicity increases with increasing Na depth for constant CO.

The deviation between input and derived metallicity has probably several contributions. The S/N caused error is small for most stars as is clear from repeated observations of the same star. At shallow CO depth, S/N can contribute as can the continuum problem of the XSL spectra. The error in the calibrator [Fe/H] is small for APOGEE (in median 0.013 and at most 0.036) and NGC6583-46 (0.08), also for the stars of \citet{Thorsbro_20} it is 0.15, still small compared to the uncertainty here. From the overlap of \citet{Arentsen_19} and APOGEE we derive an uncertainty of 0.09, consistent with their median error of 0.08. Thus, all these errors are small. That is also confirmed when we compare the scatter by spectra/metallicity calibration: all are within 1 $\sigma$ of each other when the stars with EW$_\mathrm{CO}>-2$ are excluded. 
The different data sets show somewhat different offsets, the main XSL \citep{Arentsen_19} agrees not surprisingly, since it dominates the fit. 
Surprisingly the XSL APOGEE sample has an offset of -0.2 dex with respect to the model. Since this sample consists mainly of globular clusters, this may show that they are somewhat different in these indicators. Because there are not so many nuclear disk stars in their [Fe/H] range it does not influence the results much. The SINFONI-\citet{Thorsbro_20} sample 
has an offset of -0.1 dex with respect to the model.
 That is because the mean metallicity of this sample is smaller than that of most nuclear samples, which are summarized in \citet{Schultheis_19}. Not surprisingly the APOGEE-KMOS sample 
 has an offset of 0.1 dex with respect to the model, since the model chooses a compromise between 
the APOGEE-KMOS and SINFONI-\citet{Thorsbro_20} samples which dominate the low temperature end. This difference shows that our metallicity calibration has an uncertainty of about 0.1 dex. 

\subsubsection{Application of the metallicity index}
\label{sec:met_cal_ap}

We now apply the index to our target stars. 
For checking the impact of the calibration, we also calculate [Fe/H] excluding either SINFONI-\citet{Thorsbro_20} or KMOS-APOGEE from the fit, the other samples are always used.
The indices of our spectra have errors which cause metallicity uncertainties. We calculated the resulting metallicity error by Monte Carlo (MC) simulation, i. e. we assumed a Gaussian distribution of the EW values and calculated the metallicity for 10000 random realizations. From the MC samples we calculated the 1-$\sigma$ confidence intervals. In the median it is 0.13 dex. This is small compared to the calibration error.
 It depends mostly on the S/N of the spectra, but it also matters how strong the CO and Na indices are because for strong indices, [Fe/H] varies less per \AA. For EW$_\mathrm{CO}$<-15, we find that the error is typically  $\sigma\mathrm{[Fe/H]}=0.93/\mathrm{(S/N)}$. For the smallest CO depth, the error is a factor 4 larger. Because metal poor stars  have usually shallower CO depths, 
 their random error is usually larger. We provide the errors in the result table, see Appendix~\ref{ap:star_prop}.
 It depends on the application whether this error can be used on its own or whether the calibration error should be added. The former is the case when rather similar stars are compared and only relative metallicity matters, the latter when the stars are rather different or the absolute metallicity matters.
   
   The obtained numbers confirm that the S/N cannot be the main reason for the 0.32 dex scatter in the calibration sample. A reason could be that we use the [Fe/H] of the spectra but we do not measure it. Our indices measure mainly Na and C\footnote{C is limiting in CO, see \citet{Frogel_01}.}. We looked into that possibility by using the APOGEE stars, in it 33 stars have [C/Fe] and 24 have [Na/Fe], the latter is rarely available for the most metal rich stars. For Na we do not find a trend in [Fe/H]$_\mathrm{index}-$[Fe/H]$_\mathrm{APOGEE}$; also not when we divide the sample into XSL and KMOS spectra to account for the different offsets\footnote{However, we note that Na belongs to the least reliable elements in APOGEE, see \citet{Joensson_20}.}. For [C/Fe] we find weak trends, we find a trend slope of $1.19\pm0.95$ for the APOGEE-KMOS sample and a trend slope of $0.28\pm0.26$ ($0.34\pm0.26$ excluding one outlier) for the APOGEE-XSL sample, where a slope of 1.0 would fully explain the offset. Thus, while abundance variations are probably not enough to explain the full error they likely contribute. The reason that there is no trend in [Na/Fe] could be due to the fact, that this index measures not only  Na but also Sc I for cold stars, see e.g. \citet{Park_18}.
   
\begin{figure}
   \centering
   \includegraphics[width=0.72\columnwidth,angle=-90]{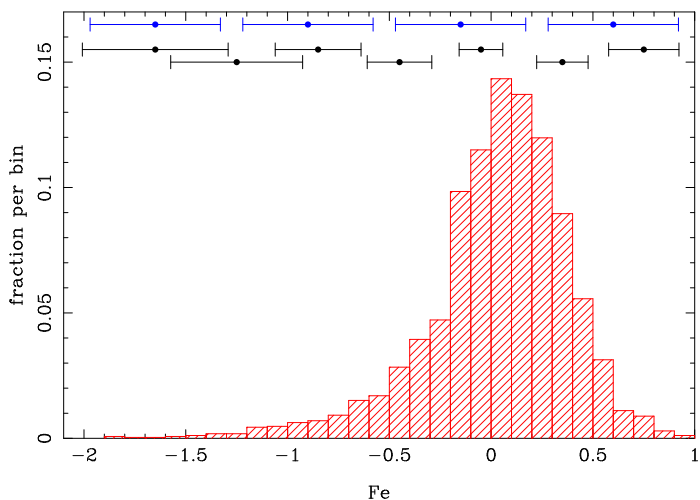}  
      \caption{Metallicity distribution of survey stars. We show all 2734 stars with CO absorption and S/N$>30$. In the top we show in black the median statistical error of stars passing these cuts and in blue the calibration uncertainty which does not depend on metallicity.
              }
         \label{fig:fe_distribution}
   \end{figure}

   We show in Figure~\ref{fig:fe_distribution} the metallicity distribution of our survey stars. 
   
 \subsection{Temperature calibration}
\label{sec:t_cal}
   
\begin{figure}
   \centering
   \includegraphics[width=0.72\columnwidth,angle=-90]{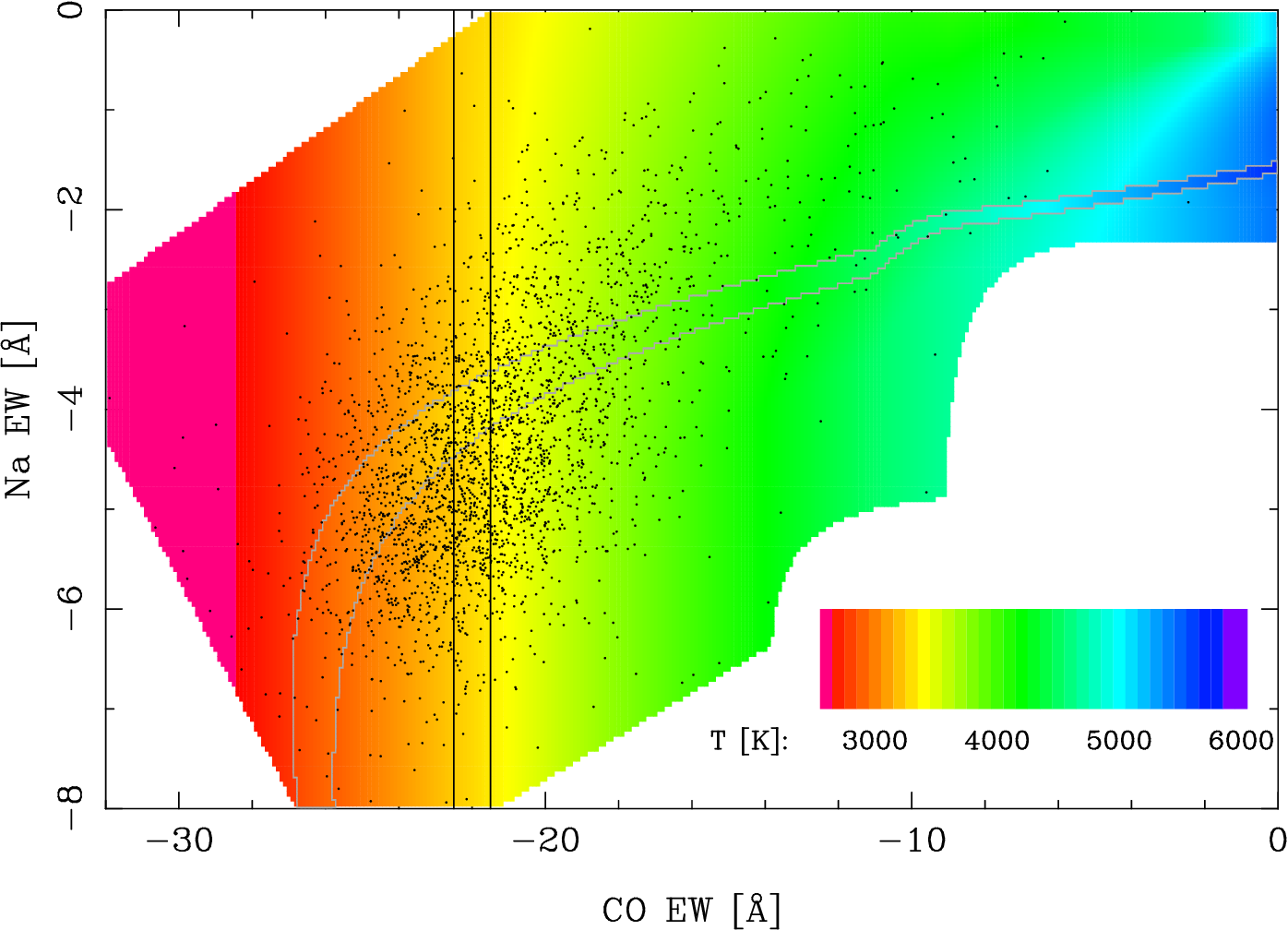}  
      \caption{Temperature calibration for stars with CO absorption. For CO EW larger than -21.5 \AA~ we use the equation of \citet{Marmol_08}. It uses our [Fe/H] determination. 
      Below -22.5 we use the calibration by \citet{Feldmeier_17a} corrected for the offset between the two. Between these values (indicated by the two black lines) we transition linearly between the two relations. The diagonal strip shows the uncorrected relation by \citet{Feldmeier_17a}. We plot it for $|[\mathrm{Fe/H]}|<0.075$. The color scales saturates outside of 2600 and 5800 K. The dots show our KMOS targets with S/N$>30$.
              }
         \label{fig:temp_calibration}
   \end{figure}

Another property which can be obtained from spectra of our resolution is the temperature.
For the cool stars which dominate our sample, a common method is to use the CO band heads. Often it is assumed that their strength depends only on the temperature, see e.g. \citet{Blum_03,Schultheis_16,Feldmeier_17a}. The strength depends however also on the metallicity, at least for metallicities clearly below solar, see e.g. \citet{Frogel_01} and \citet{Marmol_08}. To estimate the temperature we use the calibration of \citet{Marmol_08}. It uses as input our CO EW and our metallicity. 
When the metallicity is outside the range of -2.5 to 0.6 we set it to the closer limit. The range combines the range of the calibrators of \citet{Marmol_08} and of our calibrators.
For low temperatures, \citet{Marmol_08} calibrated with relatively few stars, which causes likely the following problem. Below about 3290 K the temperature increases slightly towards lower Na EW for constant CO, opposite to what is expected
 \citep{Frogel_01}.
 Therefore we use for very large CO depth the relation of \citet{Feldmeier_17a} which does not depend on metallicity. We correct for the offset between the two scales by subtracting 50 K from the \citet{Feldmeier_17a} values, that is the difference between the two scales for solar metallicity at 3290 K. We use only \citet{Marmol_08} for CO$_\mathrm{EW}>-21.5$ and only  \citet{Feldmeier_17a} adjusted for CO$_\mathrm{EW}<-22.5$. In between we transition linearly between the two. The use of \citet{Feldmeier_17a} has also the advantage that only very few stars are extremely cold. We show the temperature as function of indexes EW in Figure~\ref{fig:temp_calibration}.
  
For solar metallicity, the two scales are relatively similar, with the \citet{Marmol_08} based temperatures being usually about 100 K lower than the \citet{Feldmeier_17a} based temperatures. However, there are also differences of up to 200 K  for high temperatures\footnote{Both differences are for the unadjusted \citet{Feldmeier_17a} scale.}. 
We calculate temperature errors caused by S/N
 for all stars in a MC simulation. In that simulation we include the effects of metallicity uncertainties. The median uncertainty is 64 K for primary sources and 284 K for secondary sources. The difference is mostly caused by the higher S/N of the primary sources. Other parameters have some impact, too.
There is an error minimum around 3600 K. Those errors are usually an underestimate, the calibration of \citet{Feldmeier_17a} has a residual scatter of 163 K and the calibration of \citet{Marmol_08} of 32 K. For the calibration used, as well for other temperature calibrations \citep{Pfuhl_11,Schultheis_16,Feldmeier_17a}, we found that the resulting median temperature for solar metallicity stars is cooler than all possible temperatures from PARSEC \citep{Bressan_12} and BaSTI \citep{Pietrinferni_04} isochrones for stars with M$_{K}>-6$. This could indicate a problem with isochrones or a problem with the effective temperatures. The latter could be affected by differences between spectroscopic and photometric temperatures; spectroscopic temperatures can  be calibrated \citep{Joensson_20,Gonzalez_09} on photometric but that is difficult especially for relatively cool stars. In both cases it shows the limitations of past star formation histories obtained from spectroscopy like \citet{Blum_03,Pfuhl_11}. We give CO based temperatures for all stars with CO based velocities and negative EW$_\mathrm{CO}$ and also for stars with EW$_\mathrm{CO}<-4$ without velocity, because such a depth occurs only for late-type stars, see Figure~\ref{fig:indices}.

\begin{figure}
   \centering
   \includegraphics[width=0.72\columnwidth,angle=-90]{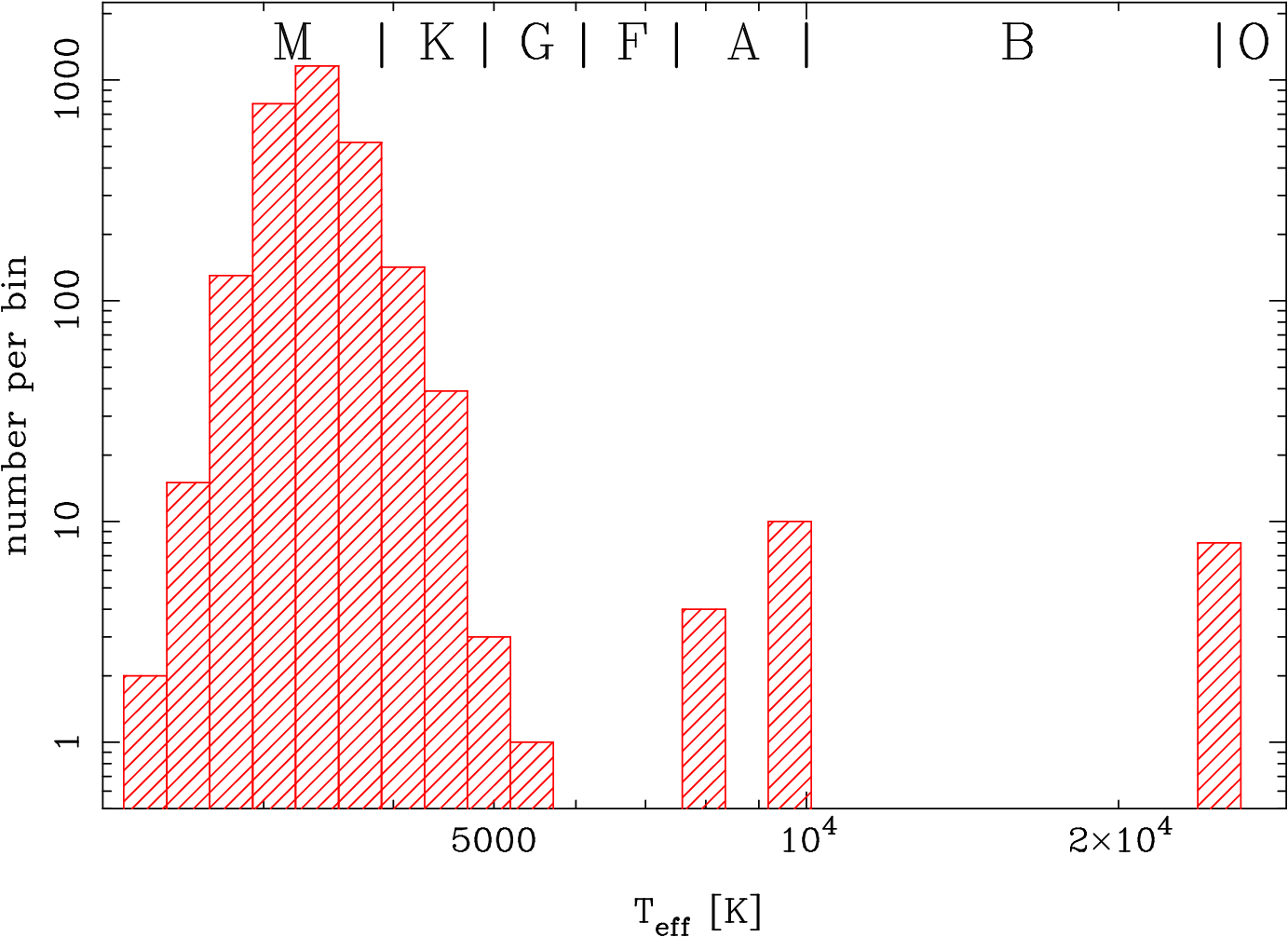}  
      \includegraphics[width=0.72\columnwidth,angle=-90]{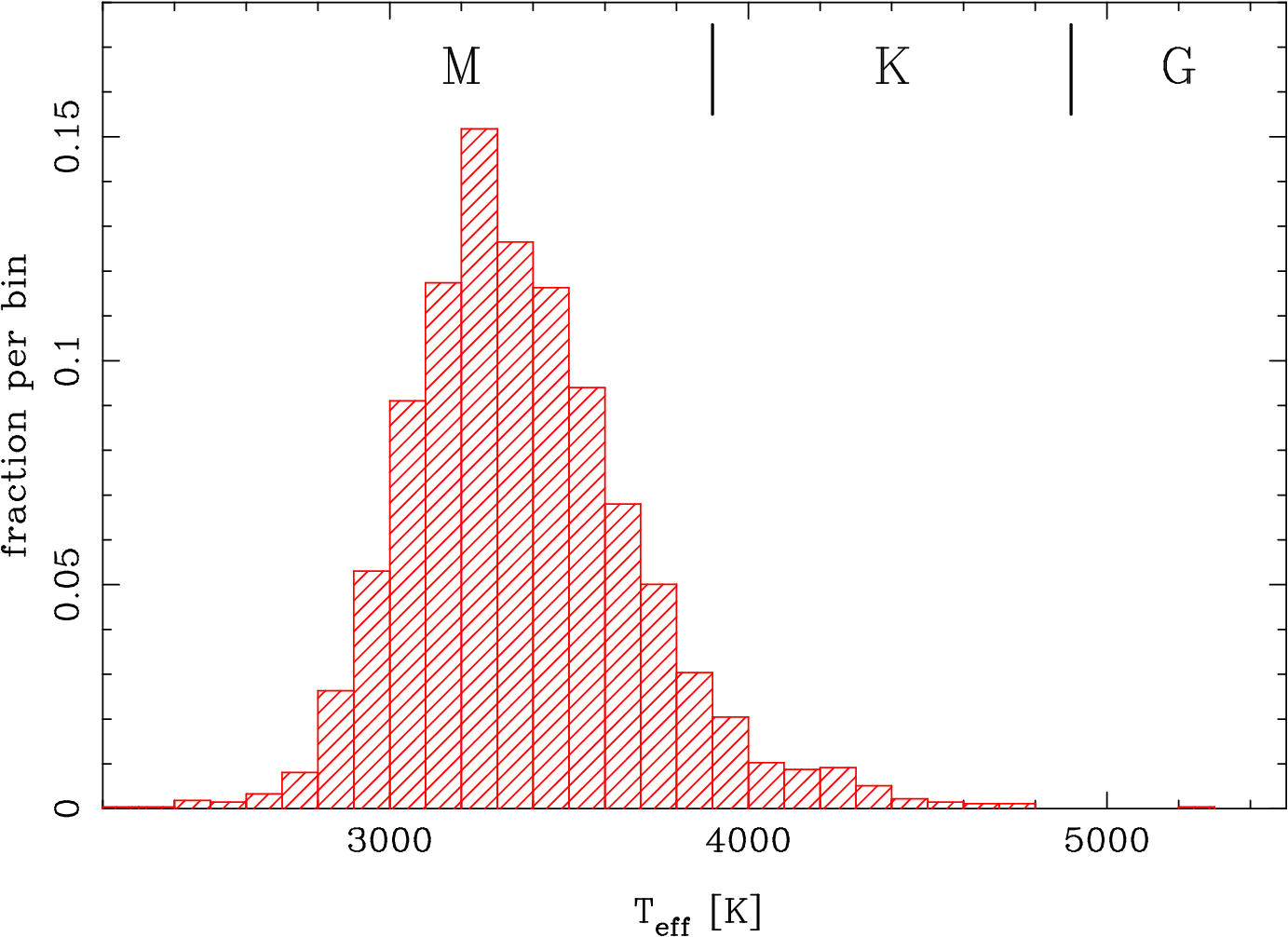} 
      \caption{Temperature distribution of survey stars. The spectral class (borders) are indicated in the top of the plots. In the top panel we show all stars with S/N$>20$, that is, all stars that we could classify (with a single exception).
       Here the latter is grouped together with the young stellar objects at 8000 K. This temperature and of the hotter stars is qualitative based on spectral feature existence. In the bottom panel we zoom into the colder stars (stars with CO absorption) for which we calculate temperatures from line indices. For the bottom panel we select stars with S/N$>30$.  
              }
         \label{fig:temp_distribution}
   \end{figure}

We cannot estimate a temperature for stars without CO absorption in this way.
 We provide rough estimates based on the lines present. 
 We assume a temperature of 10000 K for stars which show only Brackett $\gamma$ absorption and a temperature of 25000 K  for stars which show 
also or only lines with higher ionization potential.   
We assign an intermediate temperature of 8000 K to all other stars (most of them of low S/N).
We show the temperature distribution in Figure~\ref{fig:temp_distribution}. It is visible that while M-giants dominate the sample, it also contains warmer stars of different spectral classes.

We estimate from the temperatures the intrinsic H-K color for better extinction estimates. 
We use PARSEC 3.3 isochrones \citep{Bressan_12,Chen_14,Chen_15,Tang_14,Marigo_17,Pastorelli_19} for the conversion from effective temperatures to H-K colors. We round our metallicities to one decimal and set limits of -2.5 and 0.6 dex.  We download isochrones in 0.1 dex steps between -2.5 and 0.6 dex for 30 Myrs, 1 and 12 Gyrs. We use stars from M$_K=2$ (to exclude dwarfs which follow a different track) to the tip of the RGB, because later phases are less well modeled. We also exclude stars warmer than 5600 K because our stars are cooler. 
For all the selected data points we fit the H-K points with a fourth order polynomial of the temperature. The data points deviate at most by 0.014 mag in the standard deviation from these fits, likely other uncertainties such as model or passband uncertainties are more important. We extrapolate to lower temperatures by using the linear slope at the coolest data point, 21\% of the stars are in the extrapolation regime. We set stars colder than 2700 K to this temperature because no giants in PARSEC are cooler and these temperatures are likely caused by uncertainties.  
Using  the PARSEC isochrones, we assign to the hot stars (8000, 10000 and 25000 K) colors of H-K$_0=$ 0.01, -0.01 and H-K$_0$=-0.09. 
Overall our targets have an average intrinsic color H-K$_0$ of 0.30 with a scatter of 0.07; within the error consistent with our input assumption of always 0.25. 

We use these intrinsic H-K colors to order the targets in H-K corrected for the intrinsic color (H-K$_0$). All stars are compared with all potential target stars (see Section~\ref{sect:targ_select}) in that field i.e., the stars which have magnitudes between K$_0=$7 and K$_0=$9.5. For unobserved target stars, we draw the intrinsic color using the mean and standard deviation of the primary targets with S/N$>$30 in that field. Since in field 20 
most stars have low S/N, we use the average of the four neighboring fields. The extinction ordering parameter ext-order is defined as the fraction of stars which has an H-K$_0$ smaller than the target star. Ext-order orders by extinction because stellar effects are corrected for. Ext-order is mainly an estimate for the line-of-sight order, although extinction variation in the plane of sky of the fields contributes as well. Ext-order is published in our catalog, see Appendix~\ref{ap:star_prop}.

 \section{Summary and conclusions}
 \label{sec:summ}

 In this paper we introduce our spectroscopic survey of the nuclear disk. This region is highly extincted, limiting \textit{Gaia} and even APOGEE observations.
The aim of the presented survey is to study the older stellar populations in the nuclear disk region, which have not been previously systematically studied. Therefore, we target stars with absolute extinction corrected magnitudes below the tip of the red giants branch of old stars. Despite the fact that we do not use color selection criteria, most (more than 99\%) observed stars are red giants because they are the most abundant stars in the observed magnitude range. 
 
 We observe 20 fields in the nuclear disk and four reference fields in the nearby inner bulge with the multi-object IFU instrument KMOS/VLT.  We obtain K-band spectra of 3113 stars with a median S/N of 67. 
 We measure velocities for 3051 stars 
 with a typical accuracy of 5 km/s.
 We measure line indices of Brackett $\gamma$, Na, Ca, CO 2-0 and H$_2$O, in order to identify contaminants (AGB, young stars) and to measure physical properties of late-type stars.
2735 sources have sufficient S/N to 
estimate 
temperatures and metallicities which are limited by systematics.
We measure metallicities by using the two strongest features of cool ($\leq$5500 K) stars in the K-band: CO and Na. 
For calibration we use 183 giants with metallicities between -2.5 to 0.6 dex obtained with higher resolution observations and K-band spectra. 
The resulting metallicities deviate from the calibration values with a scatter of 0.32 dex. 
The internal uncertainties of our metallicities are likely smaller, since the uncertainty caused by S/N has a median value of 0.13 dex and we observe mostly similar luminousity stars. We obtain temperatures from CO using the literature relations of \citet{Marmol_08} and \citet{Feldmeier_17a}.

Our data contain also a number of early-type stars (hot stars showing hydrogen and, partly, also stars with lines with higher ionization potential and a few young stellar objects). We will publish a detailed analysis of the young stars and other rare stars in Patrick et al. in prep. They are also included in our catalog.
We publish the catalog electronically along with this paper. 
We have already analyzed (Schultheis et al. submitted) the metallicity properties and its dependency on other properties, particularly dynamics, similar to \citet{Schultheis_20} for APOGEE data covering the inner bulge and the nuclear disk. In the future we will utilize the velocities and metallicities to constrain the nuclear disk mass and dynamic state, and improve existing measurements for nuclear disk \citep{Sormani_20} and nuclear cluster \citep{Chatzopoulos_15}.
Further, we will study the star formation history of the nuclear disk (similar to \citet{Blum_03,Pfuhl_11} for the nuclear cluster), and test if the majority of stars of the nuclear disk are indeed old \citep{Nogueras_19c}.

\begin{acknowledgements}

This research made use of Astropy,\footnote{http://www.astropy.org} a community-developed core Python package for Astronomy \citep{astropy:2013, astropy:2018}.
This work has made use of data from the European Space Agency (ESA) mission
{\it Gaia} (\url{https://www.cosmos.esa.int/gaia}), processed by the {\it Gaia}
Data Processing and Analysis Consortium (DPAC,
\url{https://www.cosmos.esa.int/web/gaia/dpac/consortium}). Funding for the DPAC
has been provided by national institutions, in particular the institutions
participating in the {\it Gaia} Multilateral Agreement.
This paper includes data gathered with the 6.5 meter Magellan Telescopes located at Las Campanas Observatory, Chile.

Author R. S. acknowledges financial support from the State Agency for Research of the Spanish MCIU through the "Center of Excellence Severo Ochoa" award for the Instituto de Astrofísica de Andalucía (SEV-2017-0709). RS acknowledges financial support from national project PGC2018-095049-B-C21 (MCIU/AEI/FEDER, UE). 
N. N. and  F. N.-L. gratefully acknowledge support by the Deutsche Forschungsgemeinschaft (DFG, German Research Foundation) -- Project-ID 138713538 -- SFB 881 (`The Milky Way System', subproject B8).

\end{acknowledgements}

 \begin{appendix}
 \label{sec:app}

  \section{Details of KMOS fields}
 \label{ap:field_details}
 
In Table~\ref{tab:kmos_obs} we present the details on the observed fields and their spectra.

      \begin{table*}
      \small
      \caption[]{KMOS field observation information; a variable blue H-K cut excludes foreground stars; a field consists of 5 subsets of observations; secondary stars are sources which are found in addition to the target stars (primary sources) in the IFU observations.}
         \label{tab:kmos_obs}
      $$
         \begin{array}{p{0.11\linewidth}lllllllll}
         \hline
\hline
  \rm{Field number} &  l [\degree] & b [\degree] & \rm{blue\,} H-K \rm{\,cut} & \rm{observed\,subsets} & \rm{total\,stars} & \rm{secondary\,stars} & \rm{velocities} & \rm{stars\, with\, S/N>30}    \\
           \hline  
\hline
\hline
Nuclear disk 1 &  -0.056 &  0.043 &  0.9 &  5 &  133 &  13 &  132 &  128 \\
Nuclear disk 2 &  -0.156 &  0.173 &  0.8 &  5 &  166 &  46 &  161 &  127 \\
\hline
Bulge 3 &  -0.056 &  0.303 &  0.65 &  5 &  133 &  13 &  130 &  125 \\
Bulge 4 &  -0.056 &  0.503 &  0.5 &  5 &  126 &  8 &  126 &  119 \\
Bulge 5 &  -0.056 &  0.853 &  0.3 &  4 &  100 &  6 &  100 &  96 \\
\hline
Nuclear disk 6 &  0.104 &  0.043 &  0.9 &  0 &  0 &  0 &  0 &  0 \\
Nuclear disk 7 &  0.104 &  -0.137 &  0.9 &  4 &  120 &  1 &  97 &  97 \\
Nuclear disk 8 &  -0.216 &  0.043 &  0.8 &  0 &  0 &  0 &  0 &  0 \\
Nuclear disk 9 &  -0.216 &  -0.137 &  0.9 &  0 &  0 &  0 &  0 &  0 \\
Nuclear disk 10 &  0.264 &  -0.047 &  0.9 &  5 &  128 &  8 &  127 &  122 \\
Nuclear disk 11 &  -0.056 &  -0.137 &  0.9 &  0 &  0 &  0 &  0 &  0 \\
Nuclear disk 12 &  -0.056 &  -0.267 &  0.75 &  5 &  137 &  17 &  136 &  123 \\
Nuclear disk 13 &  0.444 &  -0.047 &  0.9 &  5 &  135 &  15 &  134 &  123 \\
Nuclear disk 14 &  0.644 &  -0.047 &  0.9 &  5 &  121 &  7 &  120 &  116 \\
Nuclear disk 15 &  0.894 &  -0.047 &  0.9 &  5 &  137 &  17 &  137 &  121 \\
Nuclear disk 16 &  1.144 &  -0.047 &  0.9 &  5 &  130 &  11 &  129 &  120 \\
Nuclear disk 17 &  1.394 &  -0.047 &  0.8 &  5 &  122 &  3 &  121 &  119 \\
Nuclear disk 18 &  -0.376 &  -0.047 &  0.9 &  5 &  132 &  12 &  125 &  113 \\
Nuclear disk 19 &  -0.556 &  -0.047 &  0.9 &  5 &  123 &  3 &  123 &  110 \\
Nuclear disk 20 &  -0.756 &  -0.047 &  0.9 &  5 &  118 &  3 &  115 &  22 \\
Nuclear disk 21 &  -1.006 &  -0.047 &  0.7 &  5 &  135 &  15 &  133 &  121 \\
Nuclear disk 22 &  -1.256 &  -0.047 &  0.9 &  5 &  123 &  3 &  123 &  121 \\
Nuclear disk 23 &  -1.506 &  -0.047 &  0.6 &  5 &  127 &  8 &  124 &  120 \\
Nuclear disk 24 &  -0.556 &  -0.247 &  0.75 &  5 &  131 &  11 &  129 &  122 \\
Nuclear disk 25 &  -0.556 &  0.153 &  0.9 &  5 &  146 &  26 &  142 &  129 \\
Nuclear disk 26 &  0.444 &  -0.247 &  0.8 &  5 &  124 &  7 &  123 &  118 \\
Nuclear disk 27 &  0.444 &  0.153 &  0.65 &  5 &  138 &  18 &  137 &  122 \\
\hline
Bulge 28 &  1.894 &  -0.047 &  0.7 &  5 &  128 &  10 &  127 &  120 \\
Bulge 29 &  -1.906 &  -0.047 &  0.65 &  0 &  0 &  0 &  0 &  0 \\
    \noalign{\smallskip}
           \hline
       \end{array}
       $$
       \normalsize
  \end{table*}

  \section{Details and tests of spectral procedures}
 \label{ap:spec_details}
 
Here we describe tests and correction affecting a minority of all spectra.
   
\subsection{Correction of sky residuals}
\label{sect:sky_res}

We checked the spectra for residual OH lines. We found that some of them, especially those extracted from IFUs without corresponding sky observations show them in emission. To quantify this effect we measure the OH flux around 2.1801, 2.1955, and 2.2125  $\mu$m. We choose these lines because they are relatively strong, in a region with less strong stellar lines and are close to the Na I doublet around 2.207 $\mu$m, which is the most important feature affected by OH emission. When at least two of these are 2 $\sigma$ above 1.0 and all three together (excluding the one which is below 2 $\sigma$) at least are 6 $\sigma$ above 1.0, we correct them for OH lines. For that we restrict the wavelength range between 2.08 and 2.29 $\mu$m, because at smaller wavelengths the telluric transmission is low and at larger wavelengths the CO absorption makes the detection of OH lines difficult. We found that there are no strong OH lines in the CO band pass. Pixels are flagged when the following three conditions are fulfilled: 
Firstly, they need to be at wavelengths of OH lines, secondly be larger than 1.07 and thirdly be at least 3 $\sigma$ larger than 1. For the two latter criteria the spectrum is normalized to 1 by a linear fit to the selected wavelength range. Those pixels are treated as bad pixels, the errors are increased to a factor 10 of the flux, the flux is corrected as for bad pixels in images. 
170 spectra are corrected in this way. In a similar way we identify and correct spectra with overcorrected OH lines. 
We find 6 such spectra. 
There are fewer because our background pixels are selected among low emission, thus we underestimate the background flux when most pixels of an IFU do not contain sources. This is good, as overcorrection of OH is more difficult to identify than undercorrection, because most of our spectra are dominated by absorption lines. 
The later mentioned young stars with emission lines are not affected by OH lines. Therefore, we use
 the original spectra for those.

 \subsection{Ripples in the KMOS Continuum}
\label{sect:spec_ripp}
 
We note that some spectra show ripples in the continuum. These are sources where a small aperture is used due to good seeing. This effect is known, see e.g. \citet{Davies_13,Feldmeier_15} and the KMOS manual. 
To quantify the affected sources we extracted for all primary sources also spectra with a circular aperture of 3 pixels. These spectra should be unaffected by the ripples but have often clearly lower S/N. We divide these spectra by our standard (potentially rippled) spectra. The result is dominated by the ripple signal. We fit the spectra ratio by a quadratic polynomial multiplied by a sine function. By visual inspection we identified the parameters characteristic for ripples, the most important one is that the sine function has at least 0.5\% amplitude. The corresponding spectra were flagged.
Ripples are mostly seen in exposures with good spatial resolution. In addition, they also populate some IFUs more than others. In particular we note, that spectra from the second spectrograph (IFU 9 to 16) show nearly no ripples.  We use the pattern in IFU and exposure to identify spectra which have likely also ripples, but whose spectra have too low S/N to identify them. In total we flag 326 primary spectra. To that we add 40 secondary sources which are on cubes where the primary source is flagged. 
While line indices of these sources are less reliable than for the other sources, especially important for Na I which dominates our metallicity estimate, visual inspection of the spectra shows that the rough metallicity and temperature of a star is still usually preserved. 
The velocities should not be affected by the ripples since a large wavelength range is used for cross-correlation and also many tests were done. 

 \subsection{Impact of no sky subtraction}
\label{sect:spec_nsky}

\begin{figure}
   \centering
   \includegraphics[width=0.72\columnwidth,angle=-90]{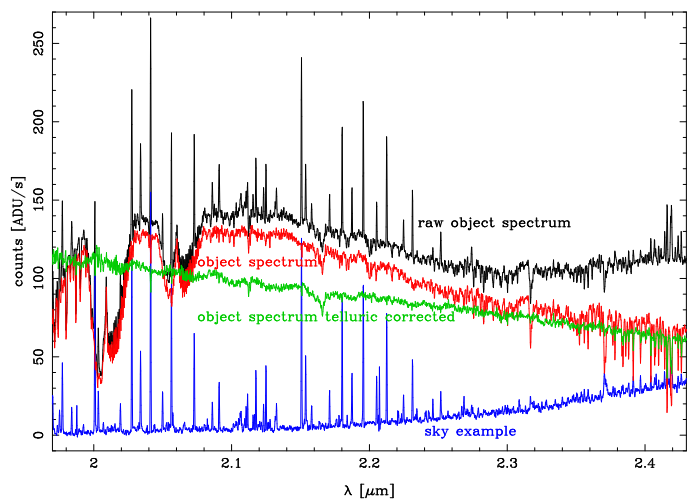}     
      \caption{Typical spectrum without sky subtraction from a separate cube. We show the raw object spectrum (black), the sky subtracted spectrum (red), a sky spectrum (blue), and the final telluric corrected science spectrum (green).
The spectrum is from 3-18-2, an early-type star (showing the line of Brackett $\gamma$ at 2.166 $\mu$m) which is a little brighter than the median magnitude of the full sample.
              }
         \label{fig:raw_spec_nsky}
   \end{figure}

For IFUs 18, 19 and 24 no
appropriate sky field could be found. As
diffuse background emission is subtracted from all IFUs, the effects of sky emission are corrected rather well, especially in the most important spectral range (2.08 to 2.32 $\mu$m), see Figure~\ref{fig:raw_spec_nsky}. Not in all cases the OH correction is as well, in those cases the in Section~\ref{sect:sky_res} explained additional correction for OH lines matters. 

To quantitatively test the impact of no dedicated sky exposure subtraction on the final spectra we now compare spectral properties of spectra with and without dedicated sky exposure subtraction. We use therefore only primary spectra, since they are more reliable. First we look at the overall properties as they come from the linear spectrum fit between 1.95 and 2.29 $\mu$m. We compare thus with the photometric properties. Firstly, we compare the zero point of the spectra with and without dedicated sky exposure subtraction, we find that in the mean and median it agrees very well with less than 0.005 mag difference. However, there is somewhat larger scatter without dedicated sky exposure subtraction
 by 12 to 22 \% dependent on whether outliers are included or not. We looked also on the other spectra obtained with the third spectrograph (That means IFU number larger than 16.) and found no increase, thus it is likely caused by the missing dedicated sky exposure subtraction. Since the overall zeropoint is not necessarily important for relevant properties, we looked also at other properties. Firstly, that is the spectral slope of the linear fit which is well correlated with the H-K color. We fit for the two cases linearly. We find that the slope is consistent for with and without dedicated sky exposure subtraction. However, the intercept differs by 0.095 (equivalent of an H-K difference of 0.11). While that is significant, it is small compared to the single star error which is 0.36. The single star error is by 11\% increased for stars with in-cube sky subtraction.

 Next we look at the velocities. We find the velocities show no difference in mean and scatter between with and without dedicated sky exposure subtraction. Most stars are in systematic error regime, but the number of the stars above the systematic error regime is 13\% higher without dedicated sky exposure subtraction than with subtraction (8\%). 
 
Now we look at the indices and derived properties from them. There are biases which are often formally significant but are always smaller than the scatter, at most 57\% of it. The bias of 0.06 is relatively large for H$_2$O, but we know that it has a systematic error of at least 0.05. Next largest bias is for Brackett $\gamma$ which is also rather unimportant. For [Fe/H] the bias is 0.06 and for T$_\mathrm{eff}$, 120 K, resulting in larger values without sky subtraction.
 The scatter increases partly dedicated sky exposure without subtraction by up to 22\% like for [Fe/H]. For T$_\mathrm{eff}$ the scatter increases by 18\%. Part of the differences with and without dedicated sky exposure subtraction are likely caused by spectrograph sub-system groups of IFUs and IFU effects.
 
There is some impact of applying only in-cube sky subtraction. However, it does not dominate the intrinsic variation. Since all fields have the missing arms, it and any IFU dependent effect does not lead to differences between fields and thus to no spurious discoveries in physical space. 
 It is possible to construct a sample that is not affected by the suboptimal in-cube only sky subtraction by excluding the affected IFUs. The second part of the identification number corresponds to the IFU.

 \section{Tests of the velocities}
\label{sec:vel_tests}

In this Appendix we test the quality of our radial velocity measurements made in Section~\ref{sec:vel_meas}. In particular, we test here whether the absolute wavelength calibration is accurate and we compare our radial velocities to the APOGEE survey.

To test whether the absolute wavelength calibration of the spectra is good enough and that its error can be considered irrelevant, we used ATRAN atmospheric transmission spectra, which we smooth to the resolution of KMOS. We tested different ATRAN cases\footnote{We tested the closest in airmass and water for  both, Cerro Panch\'{o}n and Mauna Kea, from the Gemini website and the KMOS K-model in the reduction directory.} but found that the velocities vary at most by 2 km/s (in case of the KMOS K-band model which does not vary with the conditions in contrast with the other models.), negligible compared to the other errors. 
To test for the velocity we cross-correlate the transmission spectra with the observed spectra before telluric correction of the primary targets. First, we cross-correlate the full spectra with the ATRAN models allowing offsets of up to $|600|$ km/s.  In nearly all cases cross-correlation works. We find a median velocity of -3.5 km/s with a robust scatter of 4 km/s. Since most of the telluric signal is at wavelengths very different from the spectral features, that might not be characteristic of the star velocities, if the wavelength calibration error changes with wavelength.
 For most of our stars the velocity is based on late-type spectra between 2.18 and 2.425 $\mu$m. Therefore, we also cross-correlated our spectra with transmission spectra in that range and obtain an offset of $-3.5\pm4.5$ km/s.
We also coadd spectra from the same exposure together, then the offset stays the same and the scatter reduces to about 1.5 km/s, with no trend over time. Thus the error goes down with nearly $\sqrt{N}$. This suggests that the scatter between different spectra in one exposure is probably caused by 'noise', like the spectral features in them. 
Because the CO band head in the spectra matters most for the velocity we also cross-correlate around the strongest CO features, at the strongest atmospheric feature which is at 2.317 $\mu$m. To cross-correlate only with it we use a range of 2.30 to 2.335 $\mu$m. For stars with CO lines this leads to bad results due to the many features there. Therefore we use only stars  without CO, 
the offset is $-3\pm5$ km/s.
In conclusion, the spectra are well wavelength calibrated with no evidence of wavelength distortions. There is an offset of about 4 km/s which seems rather constant, with at most 5 km/s variation over the sample.  
We correct for the calibration offset by adding 4 km/s to the raw velocities. 

\begin{figure}
   \centering
   \includegraphics[width=0.72\columnwidth,angle=-90]{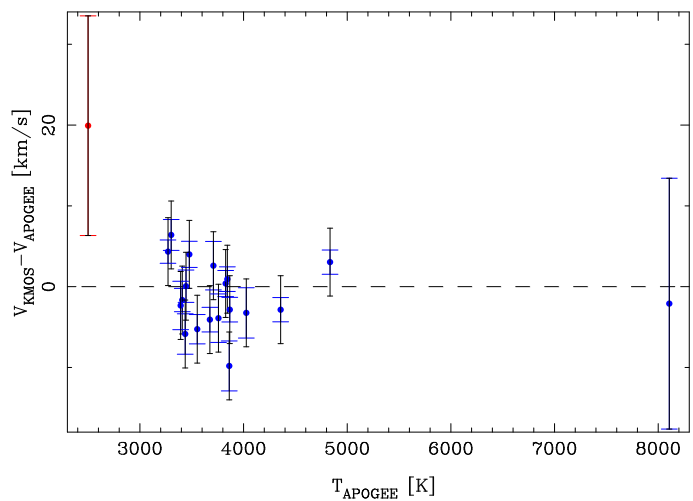}     
      \caption{Velocity offset between KMOS and APOGEE. The star in red has no effective temperature in APOGEE, it shows cold stars features. In color are shown the original KMOS errors. (The APOGEE errors are negligibly small in comparison.) The errors after adjustment are shown in black.
              }
         \label{fig:vel_offs}
   \end{figure}

In addition we compare with external velocity measurements from the APOGEE survey \citep{Majewski_17} in DR16 of SDSS \citep{Ahumada_19}. 
We find 20 common targets 
in the surveys, 19 stars with CO and one foreground star with Brackett $\gamma$ absorption. The velocities agree well in the mean, see Figure~\ref{fig:vel_offs}. However with original errors the scatter is larger than expected, as visible in the $\chi^2$ of 76. Since the agreement is better for stars with larger KMOS errors, we conclude that the KMOS errors are underestimated when the error is small. By requiring a $\chi^2$ of 19 we obtain that errors smaller than 4.2 km/s are underestimated. We therefore enlarge all errors below this limit to 4.2 km/s. The reason for the error underestimating is not clear, it could be the limit of the algorithm or calibration uncertainty. In this sample we looked at stars with several exposures and find that their error is also underestimated. Since the exposures are obtained directly after each other with the same IFUs and only small dithers, this is not surprising, because this observing strategy leads to constant systematics in the instruments. 
After the error adjustment we find that the KMOS velocities are in the weighted average -$1\pm1$ km/s smaller than the APOGEE velocities. Thus, our velocity scales are consistent. 
The enlargement of the errors does not impact our work, since the velocity differences between the survey stars are much larger than 4.2 km/s.

 \section{Metallicity calibration details}
 \label{ap:met_cal}
 
 We show in Table~\ref{tab:cal_star}, the calibrator stars used for the derivation of our metallicity relation. Besides the names and metallicity from the literature,  we also include the line indices measured by us and the derived metallicities.

   \begin{table*}
      \small
      \caption[]{Metallicity calibration stars (The full table is only available with the electronic paper.); column 1 list the name, column 2 the metallicity from the literature source (given in 3); column 4 the source of the K-band spectrum; column 5 and 6 give the line indices of Na and CO used for the mettallicity determination; column 7 gives the from us determined derived metallicity.}
         \label{tab:cal_star}
      $$
         \begin{array}{p{0.23\linewidth}lllllll}
         \hline
\hline
  name &  \mathrm{literature~[Fe/H]} & \mathrm{source~of~[Fe/H]} & \mathrm{source~of~spectrum} & \mathrm{Na~EW~[\AA]} &    \mathrm{CO~EW~[\AA]} &  \mathrm{derived~[Fe/H]}  \\
           \hline  
\hline
\hline
HD 6268                     & -2.382 & 2 & \mathrm{XSL},~3 & -0.57085 & -0.0744 & -2.01341 \\
HD 16456                    & -1.382 & 2 & \mathrm{XSL},~3 & -0.97503 & -0.11402 & -1.20498 \\
HD 139717                   & 0.298 & 2 & \mathrm{XSL},~3 & -1.37645 & -0.11079 & -0.41259 \\
HD 178287                   & 0.268 & 2 & \mathrm{XSL},~3 & -2.01691 & -0.14897 & 0.85605 \\
HD 53003                    & 0.148 & 2 & \mathrm{XSL},~3 & -1.52129 & -0.3269 & -0.16019 \\
HD 105262                   & -1.852 & 2 & \mathrm{XSL},~3 & -0.50033 & -0.35391 & -2.16084 \\
HD 179315                   & 0.338 & 2 & \mathrm{XSL},~3 & -1.77065 & -0.51751 & 0.30245 \\
HD 161817                   & -1.222 & 2 & \mathrm{XSL},~3 & -0.79191 & -0.54579 & -1.60488 \\
HD 166161                   & -1.122 & 2 & \mathrm{XSL},~3 & -0.85418 & -0.84563 & -1.50597 \\
HD 48616                    & 0.168 & 2 & \mathrm{XSL},~3 & -1.19382 & -0.90589 & -0.86107 \\
HD 2796                     & -2.442 & 2 & \mathrm{XSL},~3 & -0.6748 & -0.93804 & -1.85334 \\
HD 17072                    & -1.002 & 2 & \mathrm{XSL},~3 & -0.88558 & -1.17352 & -1.44881 \\
HD 85773                    & -2.262 & 2 & \mathrm{XSL},~3 & -0.59183 & -1.19293 & -2.01239 \\
HD 172365                   & 0.208 & 2 & \mathrm{XSL},~3 & -1.56569 & -1.2126 & -0.17291 \\
HD 186478                   & -2.222 & 2 & \mathrm{XSL},~3 & -0.8696 & -1.31837 & -1.49354 \\
HD 204543                   & -1.972 & 2 & \mathrm{XSL},~3 & -0.64096 & -1.38151 & -1.92396 \\
NGC 6397 211                & -1.972 & 2 & \mathrm{XSL},~3 & -0.8847 & -1.38269 & -1.47791 \\
BD+18 2890                  & -1.472 & 2 & \mathrm{XSL},~3 & -0.9283 & -1.38975 & -1.38489 \\
HD 165195                   & -2.182 & 2 & \mathrm{XSL},~3 & -0.17826 & -1.47843 & -2.77524 \\
NGC 7078 1079               & -2.33466 & 1 & \mathrm{XSL},~3 & -0.63271 & -1.53196 & -1.94662 \\
HD 9051                     & -1.607 & 2 & \mathrm{XSL},~3 & -1.05279 & -1.84992 & -1.19017 \\
2MASS J18352834-3444085     & -1.507 & 2 & \mathrm{XSL},~3 & -1.04436 & -1.91434 & -1.21294 \\    

    \noalign{\smallskip}
           \hline
       \end{array}
       $$
        \tablebib{
        (1)~\citet{Ahumada_19}; (2)~\citet{Arentsen_19}; (3)~\citet{Gonneau_20}; (4)~\citet{Magrini_10}; (5)~\citet{Thorsbro_20}
        }
       \normalsize
  \end{table*}

\section{Derived stellar properties}
\label{ap:star_prop}

We present the properties for all successfully observed stars in the electronic table.
A successfully observed star is defined as having a measured line-of-sight velocity and/or at least S/N$>10$.
The table lists the following properties: 
Our identification number, coordinates, magnitudes in H, K, and when available also in J  \citep[all from][]{Nishiyama_13}  
and in the IRAC bands \citep[obtained from][]{Churchwell_09,Ipac_08}. 
If there is no good match or no detection, the magnitude value is set as -999.999.
From our spectroscopy we present the overall S/N, the barycentric line-of-sight velocity and its error, the line indices of Brackett $\gamma$, Na I, Ca I, CO 2-0, H$_2$O  and their errors caused by S/N.
For all stars with a CO based temperature, we also include the metallicity, its random error term and metallicity value when we omit KMOS/APOGEE or SINFONI/\citet{Thorsbro_20} for the calibration. For all stars we include the effective temperature with its random error. For stars without CO absorption (T$>$6000 K), the temperature estimate is more uncertain as also visible in the error. For all stars we show an estimate of the intrinsic H-K color and of the extinction ordering parameter ext-order. We also provide the ripples flags. Values which could not be provided are set to -999.999.

Properties of secondary sources are often less reliable, in particular the photometric properties. 
The position is calculated from the brightest pixel offset relative to the primary source. That can be slightly offset when the maximum of either source is not within the IFU. 
The secondary sources are too faint and too confused for large scale surveys, while the deeper GALACTICNUCLEUS survey  does not cover the majority of our area.
As such we do not provide J and IRAC magnitudes for our secondary sources. We estimate K and H magnitudes from the spectra in the following way:
The K-magnitude is derived from the flux ratio comparing the secondary source relative to the primary source. Often this leads to underestimating its magnitude. We calculate the H magnitude by deriving a relationship between H-K color and K spectral slope between 1.95 and 2.29 $\mu$m for the primary sources. The scatter in this relationship is 0.37 magnitudes.
 The spectral properties can be affected by flux from the primary source.

 \end{appendix}

\bibliographystyle{aa} 
\bibliography{mspap} 
\end{document}